\documentclass[aps,eqsecnum,preprint,floats,epsf,epsfig,nofootinbib]{revtex4}
\usepackage{epsfig}
\usepackage{graphicx}

\begin{document}
\def\be{\begin{eqnarray}}
\def\en{\end{eqnarray}}
\def\non{\nonumber}
\def\la{\langle}
\def\ra{\rangle}
\def\A{{\cal A}}
\def\B{{\cal B}}
\def\c{{\cal C}}
\def\d{{\cal D}}
\def\e{{\cal E}}
\def\p{{\cal P}}
\def\t{{\cal T}}
\def\nc{N_c^{\rm eff}}
\def\CP{{\it CP}~}
\def\CPP{{\it CP}}
\def\acp{{\cal A}_{C\!P}}
\def\vp{\varepsilon}
\def\drho{\bar\rho}
\def\deta{\bar\eta}
\def\vma{{_{V-A}}}
\def\vpa{{_{V+A}}}
\def\J{{J/\psi}}
\def\ov{\overline}
\def\Lqcd{{\Lambda_{\rm QCD}}}
\def\pr{{ Phys. Rev.}~}
\def\prl{{ Phys. Rev. Lett.}~}
\def\pl{{ Phys. Lett.}~}
\def\np{{ Nucl. Phys.}~}
\def\zp{{ Z. Phys.}~}
\def\lsim{ {\ \lower-1.2pt\vbox{\hbox{\rlap{$<$}\lower5pt\vbox{\hbox{$\sim$}
}}}\ } }
\def\gsim{ {\ \lower-1.2pt\vbox{\hbox{\rlap{$>$}\lower5pt\vbox{\hbox{$\sim$}
}}}\ } }



\centerline{\large\bf Branching Fractions and {\it CP} Violation}
\vskip 0.2cm
\centerline{\large\bf in $B^-\to K^+K^-\pi^-$ and $B^-\to \pi^+\pi^-\pi^-$ Decays}
\bigskip
\centerline{\bf Hai-Yang Cheng$^{1}$, Chun-Khiang Chua$^{2}$}
\medskip
\centerline{$^1$ Institute of Physics, Academia Sinica}
\centerline{Taipei, Taiwan 115, Republic of China}
\medskip
\centerline{$^2$ Department of Physics and Center for High Energy Physics}
\centerline{Chung Yuan Christian University}
\centerline{Chung-Li, Taiwan 320, Republic of China}

\bigskip
\centerline{\bf Abstract}
\bigskip

\small
We present in this work a study of tree-dominated charmless three-body decays of $B$ mesons, $B^-\to K^+K^-\pi^-$ and $B^-\to\pi^+\pi^-\pi^-$, within the factorization approach. The main results are:
(i) There are two distinct sources of nonresonant contributions: one arises from
the $b\to u$ tree transition and the other from the nonresonant matrix element of scalar densities $\la M_1M_2|\bar q_1 q_2|0\ra^{\rm NR}$. It turns out that even for tree-dominated three-body decays, dominant nonresonant contributions originate from the penguin diagram rather than from the $b\to u$ tree process, as implied by the large nonresonant component observed recently in the $\pi^- K^+$ system which accounts for one third of the $B^-\to K^+K^-\pi^-$ rate.
(ii) The calculated branching fraction of  $B^-\to f_2(1270)\pi^-\to K^+K^-\pi^-$ is smaller than the LHCb by a factor of $\sim 7$ in its central value, but the predicted $\B(B^-\to f_2(1270)\pi^-\to\pi^+\pi^-\pi^-)$ is consistent with the data. Branching fractions of  $B^-\to f_2(1270)\pi^-$ extracted from the LHCb measurements of these two processes also differ by a factor of seven! Therefore,
it is likely that the $f_2(1270)$ contribution to $B^-\to K^+K^-\pi^-$ is largely overestimated experimentally.
Including $1/m_b$ power corrections from penguin annihilation inferred from QCD factorization (QCDF), a sizable  \CP asymmetry of 25\% in the $f_2(1270)$ component agrees with experiment.
(iii) A fraction of 5\% for the $\rho(1450)$ component in $B^-\to\pi^+\pi^-\pi^-$ is in accordance with the theoretical expectation. However, a large fraction of 30\% in $B^-\to K^+K^-\pi^-$ is entirely unexpected. This issue needs to be clarified in the future.
(iv) We study final-state  $\pi\pi\leftrightarrow K\ov K$ rescattering and find that the rescattering contributions to both $B^-\to K^+K^-\pi^-$ and $B^-\to\pi^+\pi^-\pi^-$ seem to be overestimated experimentally by a factor 4.
(v) Using the QCDF expression for the $B^-\to \sigma/f_0(500)\pi^-$ amplitude  to study the decay $B^-\to \sigma\pi^-\to\pi^+\pi^-\pi^-$, the resultant branching fraction and \CP violation of 15\%   agree with experiment.
(vi)  \CP asymmetry for the dominant quasi-two-body decay mode $B^-\to\rho^0\pi^-$ was found by the LHCb to be consistent with zero in all three $S$-wave models. In the QCDF approach, $1/m_b$ power corrections, namely, penguin annihilation and hard spectator interactions contribute destructively to $\acp(B^-\to\rho^0\pi^-)$ to render it consistent with zero.
(vii) A significant \CP asymmetry has been seen  in the $\rho^0(770)$ region for positive- and negative-helicity angle cosines. Considering the low $\pi^+\pi^-$ invariant mass region of the $B^+\to \pi^+\pi^+\pi^-$ Dalitz plot of \CP asymmetries divided into four zones, the pattern of \CP violation in each zone is well described by
the interference between $\rho(770)$ and $\sigma(500)$ as well as the nonresonant background.

\pagebreak

\section{Introduction}

In 2013 and 2014 LHCb has measured direct \CP violation in charmless three-body decays of $B$ mesons
\cite{LHCb:Kppippim,LHCb:pippippim,LHCb:2014} and found evidence of inclusive integrated \CP asymmetries $\A_{C\!P}^{\rm incl}$ in $B^+\to\pi^+\pi^+\pi^-$
(4.2$\sigma$), $B^+\to K^+K^+K^-$ (4.3$\sigma$) and $B^+\to K^+K^-\pi^+$ (5.6$\sigma$) and a 2.8$\sigma$ signal of
\CP violation in $B^+\to K^+\pi^+\pi^-$. The study of three-body decays allows to measure the distribution of \CP asymmetry in the Dalitz plot. Hence, the Dalitz-plot analysis of $\A_{C\!P}$ distributions can reveal very rich information about \CP violation. Besides the integrated \CP asymmetry, local asymmetry varies in magnitude and sign from region to region. Indeed, LHCb has also observed large asymmetries in localized regions of phase space, such as the low invariant mass region and the rescattering regions of $m_{\pi^+\pi^-}$ or $m_{K^+K^-}$ between 1.0 and 1.5 GeV.

Recently LHCb has analyzed the decay amplitudes of $B^+\to \pi^+\pi^-\pi^+$ and $B^+\to K^+K^-\pi^+$ decays in the Dalitz plot \cite{Aaij:piKK,Aaij:3pi_1,Aaij:3pi_2}. Previously, the only amplitude analysis available at $B$ factories was performed by BaBar for $B^+\to \pi^+\pi^-\pi^+$ \cite{BaBarpipipi}. In the LHCb analysis of the $B^\pm\to \pi^\pm K^+K^-$ decay amplitudes, three contributions were considered in the $\pi^\pm K^\mp$ system, namely, $K^*(892)$ and $K_0^*(1430)$ resonances plus a nonresonant contribution, and four contributions in the $K^+K^-$ system: $\rho^0(1450)$, $f_2(1270)$, $\phi(1020)$ and an amplitude accounting for the $\pi\pi\leftrightarrow K\ov K$ rescattering \cite{Aaij:piKK}.
The largest contribution with a fit fraction of 32\% comes the nonresonant amplitude in the $\pi^\pm K^\mp$ system. A surprise comes from the quasi-two-body decay $B^+\to\rho(1450)\pi^+$ which accounts for 31\% of the $K^+K^-\pi^+$ decays. This seems to imply an enormously large coupling of $\rho(1450)$ with $K^+K^-$.
Another very interesting feature of this analysis is that almost all the observed \CP asymmetry in this channel is observed in the rescattering amplitude, which is the largest \CP violation effect observed from a single amplitude.

\begin{figure}[t]
\begin{center}
\includegraphics[width=0.70\textwidth]{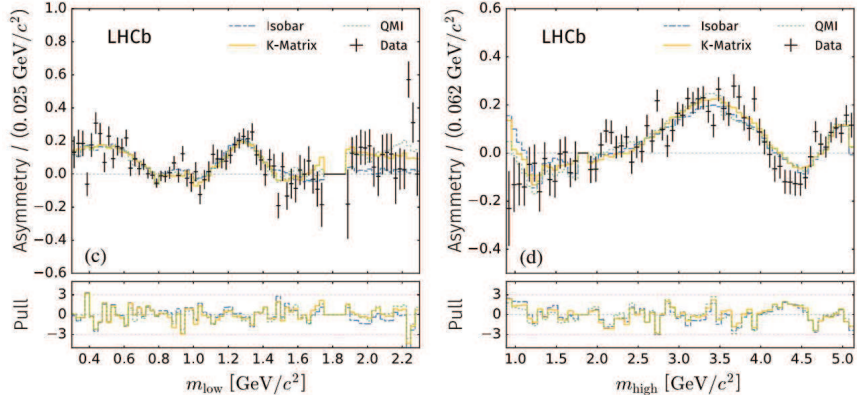}
\vspace{0.1cm}
\caption{\CP asymmetries for $B^\pm\to \pi^\pm\pi^+\pi^-$  measured in the low invariant-masss $m(\pi^+\pi^-)_{\rm low}$ region (left panel) and high
invariant-masss $m(\pi^+\pi^-)_{\rm high}$ region (right panel). This plot is taken from \cite{Aaij:3pi_2}.
}
\label{fig:Fig1}
\end{center}
\end{figure}

\begin{figure}[t]
\begin{center}
\includegraphics[width=0.70\textwidth]{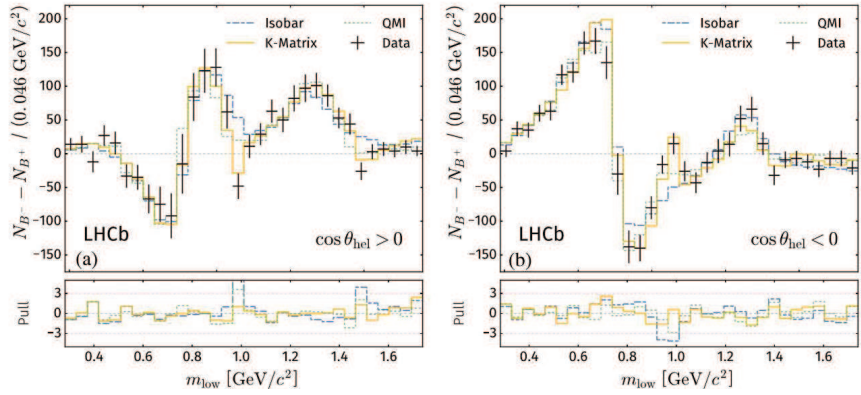}
\vspace{0.1cm}
\caption{The difference of $N_{B^-}$ and $N_{B^+}$,  the number of $B^-$ and $B^+$ events respectively, for $B^\pm\to \pi^\pm\pi^+\pi^-$ measured in the low-$m_{\rm low}$ region for (a) $\cos\theta_{\rm hel}>0$ and (b) $\cos\theta_{\rm hel}<0$ with the helicity angle $\theta_{\rm hel}$ being defined in Fig. \ref{fig:theta}. This plot is taken from \cite{Aaij:3pi_2}.}
\label{fig:Fig2}
\end{center}
\end{figure}

The LHCb analysis of the $B^-\to \pi^+\pi^-\pi^-$ decay amplitude \cite{Aaij:3pi_1,Aaij:3pi_2} showed some highlights:
(i) Instead of a large nonresonant $S$-wave  contribution observed by BaBar \cite{BaBarpipipi}, the isobar model $S$-wave amplitude was presented by the LHCb as the coherent sum of contributions from the $\sigma$ (i.e. $f_0(500)$) meson and a $\pi\pi\leftrightarrow K\ov K$ rescattering amplitude within the mass range $1.0<m_{\pi^+\pi^-}<1.5$ GeV. A significant \CP violation of  $15\%$  in $B^+\to\sigma\pi^+$ and a  large \CP asymmetry of order $45\%$ in the rescattering amplitude were found by LHCb. (ii) \CP asymmetries for $B^\pm\to \pi^\pm\pi^+\pi^-$ were measured in both low and high invariant-masss regions, see Fig. \ref{fig:Fig1}. The peak in the low-$m_{\rm low}$ region around 1.3 GeV is due to the resonance $f_2(1270)$. Indeed, the mode with $f_2(1270)$ exhibited  a \CP violation of 40\%. It is very interesting to notice a large \CP asymmetry also observed in the high-$m_{\rm high}$ region. (iii)
\CP violation in the quasi-two-body decay $B^+\to \rho^0(770)\pi^+$ is measured to be consistent with zero in all three different $S$-wave approaches, contrary to the existing model calculations. Nevertheless, a significant \CP asymmetry in the $\rho^0$ region can be seen in Fig. \ref{fig:Fig2} where the data are separated by the sign of the value of $\cos\theta_{\rm hel}$ with $\theta_{\rm hel}$ being the helicity angle, evaluated in the $\pi^+\pi^-$ rest frame, between the pion with opposite charge to the $B$ and the third pion from the $B$ decay (see Fig. \ref{fig:theta} below).
This feature which was already noticed previously in \cite{LHCb:pippippim} indicates that \CP violation close to the $\rho(770)$ resonance is proportional to $(m_\rho^2-m_{\rm low}^2)\cos\theta_{\rm hel}$. Hence, \CP asymmetry in the $\rho(770)$ region arises from the interference between the $\rho(770)$ and $S$-wave contributions. The interference pattern observed in Fig. \ref{fig:Fig2} will be destroyed by the \CP violation in $B^+\to \rho^0(770)\pi^+$ because it is proportional to $\cos^2\theta_{\rm hel}$. This is again an indication of nearly vanishing $\A_{C\!P}(B^+\to \rho^0\pi^+)$.

We have explored three-body $B$ decays in \cite{CCS:nonres,Cheng:2013dua,Cheng:2016shb} under the factorization approximation. In this work we shall update the analysis of three-body decays $B^+\to \pi^+\pi^-\pi^+$ and $B^-\to K^+K^-\pi^+$ as the LHCb has presented the new amplitude analyses of them. Attention will be paid to integrated and regional \CP violation.
We take the factorization approximation as a working hypothesis rather
than a first-principles starting point as factorization has not been proved for three-body $B$ decays.
Unlike the two-body case, to date we still do not have QCD-inspired theories for hadronic three-body decays, though attempts along the framework of pQCD and QCDF have been made in the past \cite{Wang:2014ira,Krankl:2015fha,Klein:2017xti}.

The layout of the present paper is as follows. In Sec. II we discuss the 3-body decay $B^-\to \pi^- K^+ K^-$ and take into account the intermediate state contributions from $K^*(892)$ and $K_0^*(1430)$, $\rho^0(1450)$, $f_2(1270)$ and $\phi(1020)$, a nonresonant amplitude and an amplitude accounting for the $\pi\pi\leftrightarrow K\ov K$ rescattering. In Sec. III we focus on $B^-\to \pi^- \pi^+ \pi^-$ decays. Since a clear \CP violation is seen in three places as discussed before, we shall address these three sources of \CP asymmetries. Attention is paid to the nearly vanishing \CP violation in the quasi-two-body decay $B^-\to\rho^0\pi^-$ and \CP violation induced by the interference between $S$- and $P$-wave amplitudes. Sec. IV comes to our conclusions. Input parameters for this work are summarized in Appendix A. Appendix B is devoted to the flavor operators $a_i^p$ used in this study.
Since there are some confusions in the literature concerning the final-state rescattering formula,
we shall go through the relevant derivations in Appendix C.

\section{$B^\pm\to \pi^\pm K^+ K^-$ Decays}
The charmless 3-body decays $B^-\to \pi^- K^+ K^-$ has been studied at $B$ factories by BaBar \cite{BaBarKpKmpim} and Belle \cite{BelleKpKmpim} only for its branching fraction and direct \CP asymmetry. On the theoretical side, this three-body decay mode was analyzed
in \cite{CCS:nonres,Cheng:2013dua,Cheng:2016shb} in which contributions from $K^*(892)$, $K_0^*(1430)$, $f_0(980)$ and a nonresonant amplitude were considered. The recent LHCb amplitude analysis takes into account a total of seven contributions: $K^*(892)$ and $K_0^*(1430)$, $\rho^0(1450)$, $f_2(1270)$, $\phi(1020)$, a nonresonant amplitude and an amplitude accounting for the $\pi\pi\leftrightarrow K\ov K$ rescattering. The results of the Dalitz plot analysis are shown in Table \ref{tab:Data_KKpi} \cite{Aaij:piKK}. The phases of $B^\pm$ decay amplitudes shown in the table include both weak and strong phases.
Nonresonant contributions from both $\pi^\pm K^\mp$ and $K^+K^-$ systems account for almost half of $B^\pm\to \pi^\pm K^+K^-$ rates. A very interesting feature is that the recattering amplitude, acting in the region $0.95<m_{K^+K^-}<1.42$ GeV, produced a large and negative \CP asymmetry of $(-66\pm4\pm2)\%$,
which is the largest \CP violation effect observed from a single amplitude.

\begin{table}[t]
\caption{Experimental results of the Dalitz plot fit for $B^\pm\to \pi^\pm K^+ K^-$ decays taken from \cite{Aaij:piKK}.}
\label{tab:Data_KKpi}
\begin{center}
\begin{tabular}{ l r r r r} \hline \hline
 Contribution~~~ & ~Fit fraction (\%)~~~ & ~~$\A_{C\!P}(\%)$ & ~~~$B^+$ phase ($^\circ$) & ~~$B^-$ phase ($^\circ$)\\
\hline
 $K^*(890)^0$~~ & ~$7.5\pm0.6\pm0.5$~~~ & $12.3\pm8.7\pm4.5$ & ~~$0$ (fixed)~~~ & $0$ (fixed)~~~ \\
 $K_0^*(1430)^0$~~ & ~$4.5\pm0.7\pm1.2$~~~ & $10.4\pm14.9\pm8.8$ & ~~$-176\pm10\pm16$  &  ~~$136\pm11\pm21$ \\
 NR($\pi^\pm K^\mp)$ & ~$32.3\pm1.5\pm4.1$~~~ & $-10.7\pm5.3\pm3.5$ & ~~$-138\pm7\pm5$  &  ~~$166\pm6\pm5$ \\
 $\rho(1450)^0$ & ~$30.7\pm1.2\pm0.9$~~~ & $-10.9\pm4.4\pm2.4$ & ~~$-175\pm10\pm15$  &  ~~$140\pm13\pm20$ \\
 $f_2(1270)$ & ~$7.5\pm0.8\pm0.7$~~~ & $26.7\pm10.2\pm4.8$ & ~~$-106\pm11\pm10$  &  ~~$-128\pm11\pm14$ \\
 Rescattering & ~$16.4\pm0.8\pm1.0$~~~ & $-66.4\pm3.8\pm1.9$ & ~~$-56\pm12\pm18$  &  ~~$-81\pm14\pm15$ \\
 $\phi(1020)$ & ~$0.3\pm0.1\pm0.1$~~~ & $9.8\pm43.6\pm26.6$ & ~~$-52\pm23\pm32$  &  ~~$107\pm33\pm41$ \\
\hline \hline
\end{tabular}
\end{center}
\end{table}

The explicit expression of the factorizable tree-dominated $B^-\to \pi^-(p_1)K^+(p_2)K^-(p_3)$ decay amplitude can be found in Eq. (5.1) of \cite{Cheng:2013dua}. It can be decomposed as the coherent sum of resonant contributions together with the nonresonant background
\be
A=\sum_R A_R+A_{\rm NR}.
\en
The resonant $A_R$ and nonresonant $A_{\rm NR}$ amplitudes are referred to the decay processes with and without resonant contributions, respectively. Specifically, the resonant amplitude is related to the quasi-two-body decay process which
is commonly described by the relativistic Breit-Wigner lineshape model, while the rest (at the amplitude level) is ascribed to the nonresonant contribution. In general, the nonresonant signal originates from ``direct" three-body $B$ decays.
Taking $B^-\to K^+K^-\pi^-$ as an example,  the  nonresonant contributions in our framework based on factorization arise from the nonresonant components of the 3-body matrix element $\langle K^+K^-|(V-A)_\mu|B^-\rangle$ and the 2-body matrix element of scalar density $\langle K^+\pi^-|\bar ds|0\rangle$.  Resonant contributions come from the resonant components of the above-mentioned 3-body and 2-body matrix elements. In addition,
$\langle K^+K^-|(V-A)_\mu|0\rangle$ is also governed by resonant contributions.
The presence of the nonresonant $\langle K^+\pi^-|\bar ds|0\rangle^{\rm NR}$ term induced by the penguin transition was first noticed by us together with A. Soni \cite{CCS:nonres}.

Experimentally, it is difficult to measure nonresonant contributions as the interference between the nonresonant and quasi-two-body amplitudes renders it difficult to disentangle these two distinct contributions and extract the nonresonant one.  While both BaBar and Belle have adopted the parametrization
\begin{eqnarray} \label{eq:ANR}
A_{\rm NR}=c_{12}e^{i\phi_{12}}e^{-\alpha
s_{12}^2}+c_{13}e^{i\phi_{13}}e^{-\alpha
s_{13}^2}+c_{23}e^{i\phi_{23}}e^{-\alpha s_{23}^2}
\end{eqnarray}
to describe the nonresonant three-body $B$ decays, they differ in the analysis of nonresonant
component in the $K\pi$ $S$-wave.
By contrast, LHCb did not address this issue much. In the recent LHCb analysis of $B^\pm\to \pi^+\pi^-\pi^\pm$ and $K^+K^-\pi^\pm$ decays, only the nonresonant contribution in the $\pi^\pm K^\mp$ system has been studied by the LHCb in terms of a simple single-pole form factor of the type $(1+m^2_{\pi^\pm K^\mp}/\Lambda^2)^{-1}$ \cite{Aaij:piKK}. In the experimental analysis,
it is also difficult to distinguish between the $S$-wave nonresonant background and the resonant state. For example, a large nonresonant $\pi^+\pi^-$ $S$-wave contribution observed by BaBar in $B^+\to\pi^+\pi^-\pi^+$ decays \cite{BaBarpipipi} was presented by the LHCb as the coherent sum of contributions of the $\sigma$  and a $\pi^+\pi^-\leftrightarrow K^+K^-$ rescattering amplitude.

\subsection{Resonant contributions}
In general, the intermediate vector, scalar and tensor resonances all can contribute to
the three-body matrix element $\la P_1P_2|J_\mu|B\ra$, while only the scalar resonance contributes to $\langle
P_1P_2|S|0\rangle$. Effects of intermediate resonances are described as a coherent sum of Breit-Wigner expressions. More precisely, \footnote{In \cite{Cheng:2013dua,Cheng:2016shb} an additional minus sign was wrongly put in the Breit-Wigner propagator of the scalar resonance. }
\begin{eqnarray} \label{eq:m.e.pole}
  \la K^+(p_2)K^-(p_3)|(\bar ub)_\vma|B^-\ra^R &=& \sum_i
{g^{V_i\to K^+K^-}\over
s_{23}-m_{V_i}^2+im_{V_i}\Gamma_{V_i}}\sum_{\rm
pol}\vp^*\cdot
(p_2-p_3)\la V_i|(\bar ub)_\vma|B^-\ra \non \\
&+&
 \sum_i{g^{f_{0i}\to K^+K^-}\over s_{23}- m_{f_{0i}}^2+im_{f_{0i}}\Gamma_{f_{0i}}}\langle f_{0i}|(\bar ub)_\vma|B^-\ra \non \\
&+& \sum_i {g^{f_{2i}\to K^+K^-}\over
s_{23}- m_{f_{2i}}^2+im_{f_{2i}}\Gamma_{f_{2i}}}\sum_{\rm
pol}\vp_{\mu\nu}^*p_2^\mu p_3^\nu\,\la
f_{2i}|(\bar ub)_\vma|B^-\ra, \non \\
  \langle K^+(p_2)K^-(p_3)|\bar q\gamma_\mu q|0\rangle^R &=& \sum_i
 {g^{V_i\to K^+K^-}\over s_{23}-m_{V_i}^2+im_{V_i}\Gamma_{V_i}}\sum_{\rm
pol}\vp^*\cdot(p_2-p_3)\langle V_i|\bar
q\gamma_\mu q|0\rangle, \nonumber \\
\langle \pi^-(p_1)K^+(p_2)|(\bar ds)_\vma|0\rangle^R &=&
 \sum_i {g^{K^{*0}_i\to K^+\pi^-}\over s_{12}-m_{K^{*}_i}^2+im_{K^{*}_i}\Gamma_{K^*_i}}\sum_{\rm
pol}\vp^*\cdot(p_1-p_2)\langle K^{*0}_i|
(\bar ds)_\vma|0\rangle  \non \\
&+&  \sum_i {g^{K^{*0}_{0i}\to K^+\pi^-}\over s_{12}- m_{K^*_{0i}}^2+im_{K^*_{0i}}\Gamma_{K^*_{0i}}}\langle K^{*0}_{0i}|(\bar ds)_\vma|0\rangle,  \\
 \langle K^+(p_2)K^-(p_3)|\bar dd|0\rangle^R &=&
\sum_i {g^{f_{0i}\to K^+K^-}\over s_{23}- m_{f_{0i}}^2+im_{f_{0i}}\Gamma_{f_{0i}}}\langle f_{0i}|\bar dd|0\rangle, \non \\
\langle \pi^-(p_1)K^+(p_2)|\bar ds|0\rangle^R &=&
\sum_i {g^{K^*_{0i}\to K^+\pi^-}\over s_{12}-m_{K^*_{0i}}^2+im_{K^*_{0i}}\Gamma_{K^*_{0i}}}\langle K^{*0}_{0i}|\bar
d s|0\rangle,  \non
 \end{eqnarray}
where $(\bar q_1q_2)_{V-A}=\bar q_1\gamma_\mu(1-\gamma_5)q_2$. In practice, we shall only keep the leading resonances $V_i=\phi(1020),\rho(1450)$, $f_{0i}=f_0(980)$, $f_{2i}=f_2(1270)$, $K^*_i=K^*(892)$ and $K^*_{0i}=K^*_0(1430)$.  We shall follow \cite{BSW} for the  definition of $B\to P$ and $B\to V$ transition form factors, \cite{CCH} for form factors in $B\to S$ transitions and
\cite{Wang:2010ni} for $B\to T$ transition form factors. \footnote{The $B\to T$ transition form factors defined in \cite{Wang:2010ni} and \cite{Cheng:TP} are different  by a factor of $i$. We shall use the former as they are consistent with the normalization of $B\to S$ transition given in \cite{CCH}.}

In the following we show the amplitudes from various resonances:

\vskip 0.3cm
\noindent 1. $K^*,K^*_0$:
\be
A_{K^*,K^*_0} &=&\Bigg\{ F_1^{BK}(s_{12})F_1^{K\pi}(s_{12})\left[s_{23}-s_{13}-{(m_B^2-m_K^2)(m_K^2-m_\pi^2)
\over s_{12}}\right]  \non \\
 &+& F_0^{BK}(s_{12})F_0^{K\pi}(s_{12}){(m_B^2-m_K^2)(m_K^2-m_\pi^2)\over s_{12}} \Bigg\}
 \left(a_4^p-{1\over 2}a_{10}^p\right) \non \\
 &+& { m_B^2-m_K^2\over m_b-m_s}F_0^{BK}(s_{12})m_{K^*_0}\bar f_{K^*_0} R_{K^*_0}(s_{12})\left(-2a_6^p+a_8^p\right),
\en
where
\be
&& R_{K^*_0}(s)={1\over s-m_{K^*_0}^2+i m_{K^*_0}\Gamma_{K^*_0} }, \quad F_1^{K\pi}(s)={f_{K^*}\,m_{K^*}\,g^{K^*\to K^+\pi^-}\over s-m_{K^*}^2+i m_{K^*}\Gamma_{K^*} }, \non \\
&& F_0^{K\pi}(s)=F_1^{K\pi}(s)-f_{K^*_0}g^{K^*_0\to K^+\pi^-}R_{K^*_0}(s){s\over m_K^2-m_\pi^2 }.
\en
Notice two different types of the decay constant for $K^*_0$: $f_{K^*_0}$ and $\bar f_{K^*_0}$. They are defined by $\la K^*_0(p)|(\bar d s)_\vma|0\ra=if_{K^*_0}p_\mu$ and $\la K^*_0|\bar ds|0\ra=m_{K^*_0}\bar f_{K^*_0}$, respectively.

\vskip 0.4cm
\noindent 2. $f_0(980)$
\vskip 0.2cm
It has the similar expression as the amplitude of $B^-\to\sigma/f_0(500)\pi^-\to \pi^+\pi^-\pi^-$ as will discussed in detail in the next section. Here we write down the amplitude
\be \label{eq:f0KK}
A_{f_0(980)}&=& {g^{f_0\to K^+K^-} \over s_{23}-m^2_{f_0}+im_{f_0}\Gamma_{f_0}}\Bigg\{ X^{(Bf_0,\pi)}(m_\pi^2)\left[a_1 \delta_{pu}+a^p_4+a_{10}^p-(a^p_6+a^p_8) r_\chi^\pi\right]_{f_0\pi}   \\
&+& \ov X^{(B\pi,f_0)}\left[a_2\delta_{pu} +2(a_3^p+a_5^p)+{1\over 2}(a_7^p+a_9^p)+a_4^p-{1\over 2}a_{10}^p-(a_6^p-{1\over 2}a_8^p)\bar r^{f_0}_\chi\right]_{\pi f_0}\Bigg\}, \non
\en
where
\be \label{eq:X}
X^{(Bf_0,\pi)}=-f_\pi (m_B^2-s_{23})F_0^{Bf_0^u}(m_\pi^2), \qquad
\ov X^{(B\pi,f_0)}=\bar f^d_{f_0} (m_B^2-m_\pi^2)F_0^{B\pi}(s_{23}),
\en
and
\be
r_\chi^\pi(\mu)={2m_\pi^2\over m_b(\mu)(m_u+m_d)(\mu)}, \qquad \bar r_\chi^{f_0}(\mu)={2m_{f_0}\over m_b(\mu)}.
\en
The order of the arguments of the $a_i^p(M_1M_2)$ coefficients  is dictated by the subscript $M_1M_2$ given in Eq. (\ref{eq:f0KK}).
The superscript $u$ of the form factor $F_0^{B{f_0^u}}$ reminds us that it is the $u\bar u$ quark content that gets involved in
the $B$ to $f_0$ form factor transition. Likewise, the superscript $d$ of the scalar decay constant $\bar f^d_{f_0}$ refers to the $d$ quark component of the $f_0(980)$.

\vskip 0.4cm
\noindent 3. $\phi(1020)$
\be \label{eq:phi}
A_{\phi(1020)}= -{m_\phi f_\phi g^{\phi\to K^+K^-} \over s_{23}-m^2_\phi+im_\phi\Gamma_\phi}(s_{12}-s_{13})
F_1^{B\pi}(s_{23})\bigg[a_3+a_5-\frac{1}{2}(a_7+a_9)\bigg].
\en
Since contributions from the matrix elements $\la \phi|(\bar ub)_\vma|B^-\ra$ and $\la K^+K^-|(\bar qq)_\vma|0\ra$ with $q=u,d$ to the $\phi$ production are very suppressed, their effects will not be taken into account.

\vskip 0.4cm
\noindent 4. $\rho(1450)$
\be
A_{\rho(1450)} &=& -{1\over \sqrt{2}}{g^{\rho'\to K^+K^-} \over s_{23}-m^2_{\rho'}+im_{\rho'}\Gamma_{\rho'}}(s_{12}-s_{13})\Bigg\{ {f_\pi\over 2}\Big[
2m_{\rho'}A_0^{B\rho'}(m_\pi^2)  \non \\
&+&  \left(m_B-m_{\rho'}-{m_B^2-s_{23}\over m_B+m_{\rho'}}\right)A_2^{B\rho'}(m_\pi^2)
\Big]\left[a_1 \delta_{pu}+a^p_4+a_{10}^p-(a^p_6+a^p_8) r_\chi^\pi\right]  \\
&+& m_{\rho'} f_{\rho'} F_1^{B\pi}(s_{23})\bigg[a_2\delta_{pu}
-a_4^p+\frac{3}{2}(a_7+a_9)+{1\over 2}a_{10}^p\bigg]\Bigg\}, \non
\en
with $\rho'=\rho(1450)$, where use of the relation
\be
2m_V A_3^{BV}(q^2)=(m_B+m_V)A_1^{BV}(q^2)-(m_B-m_V)A_2^{BV}(q^2)
\en
has been made.

\vskip 0.4cm
\noindent 5. $f_2(1270)$
\be \label{eq:Ampf2(1270)}
A_{f_2(1270)} &=& {1\over\sqrt{2}}{2m_{f_2}\over m_B}{f_\pi g^{f_2\to K^+K^-}\over s_{23}-m^2_{f_2}+im_{f_2}\Gamma_{f_2}}\vp^{*{\mu\nu}}(\lambda)p_{2\mu} p_{3\nu} \vp_{\alpha\beta}(\lambda)p_B^\alpha p_1^\beta A_0^{Bf_2}(m_\pi^2)
\non \\
&\times&
\left[a_1 \delta_{pu}+a^p_4+a_{10}^p-(a^p_6+a^p_8) r_\chi^\pi\right].
\en
In the approach of QCD factorization (QCDF) \cite{BBNS}, the decay amplitue of $B^-\to f_2(1270)\pi^-$ receives an additional contribution proportional to (see Eq. (B.8) of \cite{Cheng:TP})
\be \label{eq:missingf2pi}
f_{f_2}F_1^{B\pi}(m_{f_2}^2)\left[a_2\delta_{pu}+2(a_3^p+a_5^p)+a_4^p+r_\chi^{f_2}a_6^p+{1\over 2}(a_7^p+a_9^p)-{1\over 2}(a_{10}^p+r_\chi^{f_2}a_8^p)\right].
\en
The reader is referred to \cite{Cheng:TP} for the definition of the decay constant $f_{f_2}$ and the chiral factor $r_\chi^{f_2}$.
As stressed in \cite{Cheng:TP},  the factorizable amplitude $\la f_2 |J^{\mu}|0\ra\la \pi^-|J'_{\mu}|B^- \ra$ vanishes in the factorization approach as the tensor meson cannot be produced through the $V-A$ or tensor current. Nevertheless, beyond the factorization approximation, contributions proportional to the decay constant $f_{f_2}$ can be produced from vertex, penguin and spectator-scattering corrections.

Using the relation
\begin{eqnarray} \label{eq:polarization}
 \sum_\lambda \epsilon_{\mu\nu}(\lambda)\epsilon^\ast_{\rho\sigma}(\lambda)
  = \frac12 M_{\mu\rho} M_{\nu\sigma}+\frac12 M_{\mu\sigma} M_{\nu\rho}
  -\frac13 M_{\mu\nu} M_{\rho\sigma}\,,
\end{eqnarray}
with $M^{\mu\nu} = g^{\mu\nu} - P^\mu P^\nu/m_{f_2}^2$ and $P=p_2+p_3$, it is straightforward to show that \cite{Dedonder:2010fg}
\be \label{eq:tensorangular}
\sum_\lambda \vp^{*{\mu\nu}}(\lambda)\vp_{\alpha\beta}(\lambda)p_{2\mu} p_{3\nu} p_B^\alpha p_1^\beta={1\over 3}(|\vec{p}_1||\vec{p}_2|)^2-(\vec{p}_1\cdot\vec{p}_2)^2,
\en
with
\be \label{eq:3momentum}
|\vec{p}_1|=\left({(m_B^2-m_\pi^2-s_{23})^2\over 4s_{23}} - m_\pi^2\right)^{1/2}, \quad |\vec{p}_2|=|\vec{p}_3|={1\over 2}\sqrt{s_{23}-4m_K^2},
\en
and
\be \label{eq:angle}
\vec{p}_1\cdot \vec{p}_2={1\over 4}(s_{13}-s_{12}),
\en
where $\vec{p}_1$ and $\vec{p}_2$ are the momenta of the $\pi^-(p_1)$ and $K^+(p_2)$, respectively, measured in the rest frame of the dikaon $K^+(p_2)$ and $K^-(p_3)$.
However, the predicted \CP asymmetry is of order $-0.01$ which is wrong in sign and magnitude compared to experiment, especially a large \CP violation of 40\% observed in the decay  $B^-\to \pi^- f_2(1270)\to\pi^-\pi^+\pi^-$. We thus follow the QCDF calculation in \cite{Cheng:TP} to include $1/m_b$ power corrections arising from penguin annihilation (see Eq. (B.8) in \cite{Cheng:TP}). This amounts to adding the penguin annihilation contributions  $\beta_2^p\delta_{pu}+\beta_3^p+\beta^p_{\rm 3,EW}$ to the $[\dots]$ term in Eq. (\ref{eq:Ampf2(1270)}). Therefore,  the amplitude $A_{f_2(1270)}$ reads
\be
A_{f_2(1270)} &=& \sqrt{2}\,{m_{f_2}\over m_B}{f_\pi g^{f_2\to K^+K^-}\over s_{23}-m^2_{f_2}+im_{f_2}\Gamma_{f_2}} A_0^{Bf_2}(m_\pi^2)\left[{1\over 3}(|\vec{p}_1||\vec{p}_2|)^2-(\vec{p}_1\cdot\vec{p}_2)^2\right]
\non \\
&\times&
\left[a_1 \delta_{pu}+a^p_4+a_{10}^p-(a^p_6+a^p_8) r_\chi^\pi +\beta_2^p \delta_{pu}+\beta_3^p+\beta^p_{\rm 3,EW}\right].
\en
Numerically, we shall follow \cite{Cheng:TP} to use
\be
\beta_2^p(f_2\pi)=0.023-i0.011, \qquad (\beta_3^p+\beta_{\rm 3,EW}^p)(f_2\pi)=-0.047+i 0.053\,.
\en

It should be remarked that the angular momentum distribution for the vector or tensor intermediate state is not put by hand. It will come out automatically in the factorization approach. For example, the decay amplitude of $\rho(1450)$ or $\phi$ production contains a term $(s_{12}-s_{13})$ which is proportional to $\vec{p}_1\cdot\vec{p}_2=|p_1||p_2|\cos\theta_{12}$ (see Eq. (\ref{eq:angle})). Likewise, the angular distribution of a tensor meson decaying into two spin-zero particles is governed by  $(3\cos^2\theta_{12}-1)$ [cf. Eq. (\ref{eq:tensorangular})]. In general, the angular momentum distribution is described by the Legendre polynomial $P_J(\cos\theta)$.

\subsection{Nonresonant contributions}
The nonresonant contributions arise from the 3-body matrix element $\la K^+(p_2)K^-(p_3)|(\bar ub)_\vma|B^-\ra^{\rm NR}$ in the $K^+K^-$ system and the 2-body matrix element of scalar density $\langle \pi^-(p_1)K^+(p_2)|\bar ds|0\rangle^{\rm NR}$ in the $\pi^-K^+$ system.
The nonresonant contribution to the three-body matrix element can be parameterized in terms of four unknown form factors. The general expression of the nonresonant amplitude in the $K^+K^-$ system induced from the $b\to u$ tree transition reads
\be \label{eq:AHMChPT}
 A_{\rm NR}^{\rm HMChPT} &\equiv& \la K^+ (p_2) K^-(p_3)|(\bar u b)_{V-A}|B^-\ra^{\rm NR}\la \pi^-(p_1)|(\bar d u)_{V-A}|0\ra  \non\\
 &=& -\frac{f_\pi}{2}\left[2 m_\pi^2 r+(m_B^2-s_{23}-m_\pi^2) \omega_+
 +(s_{12}-s_{13}) \omega_-\right],
\en
where the form factors $r$ and $\omega_\pm$ can be calculated using heavy meson chiral perturbation theory (HMChPT) \cite{LLW,Fajfer:1998yc}. However, HMChPT is applicable only when the two scalars $K^+$ and $K^-$ in $B\to K^+K^-$ transition are soft. Indeed, the predicted nonresonant rate, of order $33\times 10^{-6}$ in branching fraction, based on HMChPT will be one order of magnitude larger than the world average of the total branching fraction $\sim 5.2\times 10^{-6}$.
Hence, we shall assume the momentum dependence of nonresonant amplitudes in an exponential form \cite{CCS:nonres}
\be \label{eq:ADalitz}
  A_{\rm NR}^{K^+K^-}=A_{\rm NR}^{\rm
  HMChPT}\,e^{-\alpha_{_{\rm NR}} p_B\cdot(p_2+p_3)}\left[a_1 \delta_{pu}+a^p_4+a_{10}^p-(a^p_6+a^p_8) r_\chi^\pi\right],
\en
in analog to Eq. (\ref{eq:ANR}),
so that the HMChPT results are recovered in the soft meson limit $p_2,~p_3\to 0$.
For the parameter $\alpha_{_{\rm NR}}$  we shall use $\alpha_{_{\rm NR}}=0.160\,{\rm GeV}^{-2}$.
\footnote{The parameter $\alpha_{_{\rm NR}}=0.081^{+0.015}_{-0.009}\,{\rm GeV}^{-2}$ used in
\cite{Cheng:2013dua,Cheng:2016shb} was originally constrained from the BaBar's measurement of the nonresonant contribution to $B^-\to \pi^+\pi^-\pi^-$  \cite{BaBarpipipi}. However, a substantial part of the nonresonant amplitude is now replaced by the scalar $\sigma$ meson in the LHCb analysis based on the isobar model. This leads to a larger $\alpha_{_{\rm NR}}$.}

The extrapolation from the soft meson limit where HMChPT is applicable to the physical kinematic region through Eq. (\ref{eq:ADalitz}) is our main ansatz for the tree nonresonant amplitude. In the literature, similar $B\to \pi\pi$ form factors in $B\to \pi\pi\pi$ decays have been studied extensively \cite{Faller:2013dwa,Kang:2013jaa,Hambrock:2015aor,Boer:2016iez,Cheng:2017smj,Cheng:2017sfk,Cheng:2019hpq}.
In the kinematic regime where the dipion state has a large energy and a low invariant mass, $\bar B^0\to \pi^+\pi^0$ form factors have been studied using QCD light-cone sum rules with $B$-meson distribution amplitudes \cite{Cheng:2017smj,Cheng:2017sfk,Cheng:2019hpq}. The interference of $\rho(770)$ with the excited $\rho$ resonances such as $\rho(1450)$ and $\rho(1750)$ are regarded as effective ``nonresonant" contributions. At low recoil (low $q^2$) and small dipion invariant mass, form factors for the pion-pion system have been treated in dispersion theory \cite{Kang:2013jaa}. One can match dispersion theory with HMChPT to fix the subtraction constant and access the low-energy-region physics. In this work we will not use the results from light-cone sum rules or dispersion theory as the nonresonant contributions there are not specified separately and explicitly. It is worth mentioning that form factors at large dipion invariant masses can be calculated in QCD factorization \cite{Boer:2016iez}. Neglecting possible resonant effects at large $\pi\pi$ invariant mass, QCD factorization provides a direct calculation of nonresonant contributions.

The nonresonant contribution in the  $\pi^-K^+$ system is given by
\be
A_{\rm NR}^{\pi^-K^+}&=&  \la K^-(p_3)|\bar s b|B^-\ra \la \pi^-(p_1)K^+(p_2)|\bar d s|0\ra^{\rm NR}(-2 a^p_6+a^p_8) \non \\
       &=& {m_B^2-m_K^2\over m_b-m_s}F_0^{BK}(s_{12})\la \pi^-(p_1)K^+(p_2)|\bar d s|0\ra^{\rm NR}(-2 a^p_6+a^p_8),
\en
where the nonresonant matrix element of scalar density has the expression \cite{Cheng:2013dua,Cheng:2016shb}
\be \label{eq:KpimeNew2}
 \la \pi^-(p_1)K^+(p_2)|\bar d s|0\ra^{\rm NR}
 = \sigma_{_{\rm NR}}
 e^{-\alpha s_{12}}\left(1-4{m_K^2-m_\pi^2\over s_{12}}\right),
\en
with \footnote{The value of the parameter
$\sigma_{_{\rm NR}}$  given in \cite{CCS:nonres} was determined at the scale $\mu=\ov m_b/2$. In this work, we will confine ourselves to the renormalization scale $\mu=\ov m_b(\ov m_b)=4.18$ GeV, see also Appendix B. In our previous work \cite{Cheng:2013dua,Cheng:2016shb} we employed the BaBar measurement $\alpha=(0.14\pm0.02){\rm GeV}^{-2}$ \cite{BaBarKpKmK0}. In order to fit to the nonresonant rate in the $\pi^-K^+$ system, the value of $\alpha$ is reduced by a factor of 2.}
\be \label{eq:sigma}
  \sigma_{_{\rm NR}}= e^{i\phi_{\pi\!K}}\left(4.74^{+0.25}_{-0.29}\right)\,{\rm GeV}, \qquad
  \alpha=0.069\,{\rm GeV}^{-2},
\en
where the phase $\phi_{\pi K}$ will be specified later.
As stressed in \cite{Cheng:2016shb}, the nonresonant signal in the $\pi^-K^+$ system is governed by the nonresonant component of the matrix element of scalar density.
Owing to the exponential suppression factor $e^{-\alpha s_{12}}$, the nonresonant contribution manifests in the low invariant mass regions.
Note that in the LHCb analysis, the nonresonant amplitude is parameterized in terms of a simple single-pole form factor of the type $(1+m^2_{\pi^\pm K^\mp}/\Lambda^2)^{-1}$ with $\Lambda\sim 1$ GeV. We prefer to use the exponential form for nonresonant amplitudes.

\subsection{Final-state rescattering}
\CP asymmetries (integrated or regional) measured by the LHCb are positive for $h^-\pi^+\pi^-$ and negative for $h^-K^+K^-$ with $h=\pi$ or $K$. The former usually has a larger \CP asymmetry in magnitude than the latter. This has led to the conjecture that $\pi^+\pi^-\leftrightarrow K^+K^-$
rescattering may play an important role in the generation of the strong phase
difference needed for such a violation to occur \cite{LHCb:2014}. The $C\!PT$ theorem requires
that $\Delta\Gamma_\lambda^{\rm FSI} \equiv \Gamma(B\to\lambda)^{\rm FSI}-\Gamma(\bar B\to \bar\lambda)^{\rm FSI}$ be vanished when summing over all the possible states allowed by final-state interactions; that is, $\sum_\lambda\Delta \Gamma_\lambda^{\rm FSI}=0$. However, in the LHCb analysis, only the two channels $\alpha=\pi^+\pi^-P^-$ and $\beta=K^+K^-P^-$ ($P=\pi,K$) in $B^-$ decays are assumed to be strongly coupled through final-state interactions with the third meson $P$ being treated as a bachelor or a spectator. It follows that $\Delta\Gamma^{\rm FSI}_\beta=-\Delta\Gamma^{\rm FSI}_\alpha$. It was found that final-state rescattering of $\pi^+\pi^-\leftrightarrow K^+K^-$ dominates the asymmetry in the mass region between 1 and 1.5 GeV. In reality, the consideration of only rescattering between $\pi^+\pi^-$ and $K^+K^-$ in the $S$-wave configuration is too restrictive and simplified \cite{Hou:2019xfk}.
For example, $\pi^+\pi^-$ is allowed to rescatter into $K^+K^-$ with charge neutral multi-pion states. Nevertheless, below we shall follow the work of \cite{Pelaez:2004vs} (also the same framework adapted in \cite{Bediaga:2015}) to describe the inelastic $\pi\pi\leftrightarrow K\ov K$ rescattering process and consider this final-state rescattering effect on inclusive and local \CP violation.

Neglecting possible interactions with the third meson under the so-called `2$+$1' assumption,
the $S$-wave $\pi^+\pi^-\leftrightarrow K^+K^-$ rescattering through final-state interactions is described by \cite{Chua:2007cm,Suzuki} \footnote{This is different from our previous treatment in \cite{Cheng:2016shb} in one aspect, namely, only the $S$-wave $\pi\pi$ and $K\ov K$ amplitudes will undergo final-state rescattering.}
\be \label{eq:FSI}
\left(\begin{array}{c} A(B^-\to \pi^+\pi^-P^-) \\ A(B^-\to K^+K^-P^-)
\end{array}\right)^{\rm FSI}_{\rm S-wave}=S^{1/2} \left(\begin{array}{c} A(B^-\to \pi^+\pi^-P^-)\\ A(B^-\to K^+K^-P^-)
\end{array}\right)_{\rm S-wave}
\en
with $P=\pi,K$. The unitary $S$ matrix reads
\be
S=\left(
\begin{array}{cc}
\eta e^{2i\delta_{\pi\pi}} & i\sqrt{1-\eta^2} e^{i(\delta_{\pi\pi}+\delta_{K\!\bar K})} \\
i\sqrt{1-\eta^2} e^{i(\delta_{\pi\pi}+\delta_{K\!\bar K})} & \eta e^{2i\delta_{K\!\bar K}}
\end{array}
\right),
\en
where the inelasticity parameter $\eta(s)$ is given by \cite{Pelaez:2004vs}
\be
\eta(s)=1-\left(\epsilon_1{k_2\over s^{1/2} }+\epsilon_2{k_2^2\over s}\right)\,{ {M'}^2-s\over s},
\en
with
\be
k_2={\sqrt{s-4m_{K(\pi)}^2}\over 2}
\en
for rescattering to a pair of kaons (pions).
The $\pi\pi$ phase shift has the expression
\be
\delta_{\pi\pi}(s)={1\over 2}\cos^{-1}\left( {\cot^2[\delta_{\pi\pi}(s)]-1\over \cot^2[\delta_{\pi\pi}(s)]+1 }\right),
\en
with
\be
\cot[\delta_{\pi\pi}(s)]=c_0\,{(s-M_s^2)(M_f^2-s)\over M_f^2 s^{1/2} }\,{|k_2|\over k_2^2}.
\en
We shall assume that $\delta_{K\!\bar K}\approx \delta_{\pi\pi}$ in the rescattering region.
We have shown in \cite{Cheng:2016shb} that the matrix $S^{1/2}$ can be expressed as
\be \label{eq:S1/2}
S^{1/2}= e^{i\delta_{\pi\pi}} \left(
\begin{array}{cc}
\cos\phi/2 & i\sin\phi/2  \\
i\sin\phi/2 & \cos\phi/2
\end{array}
\right),
\en
with
\be
\phi=\tan^{-1}{\sqrt{1-\eta^2}\over \eta}.
\en
For numerical calculations we shall use the parameters given in Eqs.~(2.15b') and (2.16) of \cite{Pelaez:2004vs}, namely $M'=1.5$ GeV, $M_s=0.92$ GeV, $M_f=1.32$ GeV, $\epsilon_1=2.4$, $\epsilon_2=-5.5$ and $c_0=1.3$\,.

The rescattering amplitude reads from Eqs. (\ref{eq:FSI}) and (\ref{eq:S1/2}) to be
\be \label{eq:Ampres}
A(B^-\to K^+K^-\pi^-)_{\rm rescattering} &=& e^{i\delta_{\pi\pi}}\Big[\cos(\phi/2)A(B^-\to K^+K^-\pi^-)_{\rm S-wave}  \non \\
&+& i\sin(\phi/2)A(B^-\to
\pi^+\pi^- \pi^-)_{\rm S-wave}\Big].
\en
The $S$-wave amplitudes involved in rescattering are given by
\be \label{eq:Swave}
A(B^-\to K^+K^-\pi^-)_{\rm S-wave} &=& A_{\rm NR}^{K^+K^-}+A_{f_0(980)}, \non \\
A(B^-\to \pi^+\pi^-\pi^-)_{\rm S-wave} &=& A_{\rm NR}^{\pi^+\pi^-}+A_{\sigma(500)}.
\en
The nonresonant amplitude $A_{\rm NR}^{\pi^+\pi^-}$ and the amplitude with the scalar resonance $\sigma(500)$ will be discussed in Sec.III.

Eq. (\ref{eq:FSI}) is sometimes expressed in the literature in terms of the $S$ matrix  instead of $S^{1/2}$. For example, writing the decay amplitude as ${\cal A}^\pm=A_\lambda+B_\lambda e^{\pm i\gamma}$, it has been shown in \cite{Bediaga:2015} that
the lowest-order (LO) effect due to FSI in the decay amplitude is given by (see Eq. (18) of \cite{Bediaga:2015})
\be \label{eq:AmpS}
{\cal A}^\pm_{\rm LO}=A_{0\lambda}+e^{\pm i\gamma}B_{0\lambda}+i\sum_{\lambda'}t_{\lambda',\lambda}(A_{0\lambda'}+e^{\pm i\gamma}B_{0\lambda'})\to \sum_{\lambda'}S_{\lambda,\lambda'}(A_{0\lambda'}+e^{\pm i\gamma}B_{0\lambda'}),
\en
where use of $S_{ij}=S_{ji}$ has been made.
However, the reader can check that the above amplitude does not satisfy Eq. (12) of  \cite{Bediaga:2015} up to the leading order in $t_{\lambda',\lambda}$, namely,
\be
A_\lambda+e^{\mp i\gamma}B_\lambda=\chi_h\chi_\lambda(A_\lambda+e^{\pm i\gamma}B_\lambda)^*+
i\chi_h\chi_\lambda\sum_{\lambda'}t_{\lambda',\lambda}(A_{\lambda'}+e^{\pm i\gamma}B_{\lambda'})^*,
\en
with the relations $A_{0\lambda}=\chi_h\chi_\lambda A_{0\lambda}^*$ and $B_{0\lambda}=\chi_h\chi_\lambda B_{0\lambda}^*$.
The correct answer should read
\be \label{eq:AmpS1/2}
{\cal A}^\pm_{\rm LO}=A_{0\lambda}+e^{\pm i\gamma}B_{0\lambda}+{i\over 2}\sum_{\lambda'}t_{\lambda,\lambda'}(A_{0\lambda'}+e^{\pm i\gamma}B_{0\lambda'})\to
\sum_{\lambda'}S^{1/2}_{\lambda,\lambda'}(A_{0\lambda'}+e^{\pm i\gamma}B_{0\lambda'}).
\en
Hence, Eq. (\ref{eq:FSI}) gives the correct description of $\pi^+\pi^-\leftrightarrow K^+K^-$ final-state rescattering.
\footnote{
In Eq. (\ref{eq:FSI}) we have used the factorized amplitude ${\cal A}_{\rm fac}$ in the place of $A_{0\lambda'}+e^{\pm i\gamma}B_{0\lambda'}$. They are, however, not exactly the same.
In fact, we are using a time evolution picture~\cite{tdlee, wshou} and the rescattering of $\pi\pi\to K\ov K$ happens at a much later stage of time-evolution.
The full amplitude should read
${\cal A}={\cal S}^{1/2} {\cal A}_0$ with ${\cal A}_0$ being free from any strong phase, and the $S$-matrix ${\cal S}^{1/2}$ corresponds to a time-evolution operator $U(\infty,0)$~\cite{tdlee} (see Appendix C for details).
Then we separate the time-evolution operator into $U(\infty,0)=U(\infty,\tau)U(\tau,0)$ with $\tau$ being short enough to treat quarks and gluons as good degrees of freedom. Consequently, the strong phase in $U(\tau,0) {\cal A}_0$ can be calculated in the factorization approach giving ${\cal A}_{\rm fac}=U(\tau,0) {\cal A}_0$~\cite{wshou,Chua:2007cm}.
Hence, the full amplitude becomes ${\cal A}=U(\infty,\tau) {\cal A}_{\rm fac}$, which corresponds to Eq. (\ref{eq:FSI}) with $\pi\pi\to K\ov K$ rescattering contained in $U(\infty,\tau)=S^{1/2}$.
}
However, the LHCb analysis of $\pi\pi\leftrightarrow K\ov K$ rescattering is based on the model described in \cite{Bediaga:2015,Bediaga:2013ela}.

\subsection{Numerical results and discussions}
The total decay amplitude of $B^-\to \pi^-K^+K^-$ now reads
\be
A(B^-\to \pi^-K^+K^-)&=&{G_F\over\sqrt{2}}\sum_{p=u,c}\lambda_p\Big(A_{K^*(892)}+A_{K^*_0(1430)}
+A_{f_0(980)}+A_{\phi(1020)}
+A_{\rho(1450)}  \non \\
&+& A_{f_2(1270)} + A_{\rm NR}^{\pi^-K^+} + A_{\rm NR}^{K^+K^-}+A_{\rm rescattering}\Big),
\en
with $\lambda_p\equiv V_{pb} V^*_{pd}$.

The strong coupling constants such as $g^{K^{*0}\to\pi^-K^+}$ and
$g^{f_0(980)\to K^+K^-}$,$\cdots$ etc., are determined from the measured
partial widths through the relations \footnote{There is some confusion in the literature concerned with the relation between the width and coupling for the tensor meson. For example, it was expressed as
$\Gamma_{T\to P_1P_2}=\alpha_{_{T P_1P_2}}{p_c^5\over 60\pi m_T^2}g_{T\to P_1P_2}^2$ \cite{Giacosa:2005bw},
where the factor of  $\alpha_{_{T P_1P_2}}$ takes into account the average over spin of the initial state and sum over final isospin states with averaging over initial isospin states, while the relation
 $\Gamma_{T\to P_1P_2}={p_c^5\over 15\pi m_T^2}g_{T\to P_1P_2}^2$
 was used in \cite{Suzuki:1993} and \cite{Pilkuhn}. It turns out that the narrow width approximation, for instance, $\Gamma(B^+\to f_2 \pi^+\to \pi^+\pi^-\pi^+)=\Gamma(B^+\to f_2 \pi^+)\B(f_2\to \pi^+\pi^-)$ is respected if the tensor couping and width satisfy the relation given in Eq. (\ref{eq:partialwidth_1}).
 }
 \be \label{eq:partialwidth_1}
 && \Gamma_{S\to P_1P_2}={p_c\over 8\pi m_S^2}g_{S\to P_1P_2}^2,\quad
 \Gamma_{V\to P_1P_2}={p_c^3\over 6\pi m_V^2}g_{V\to P_1P_2}^2,\quad
 \Gamma_{T\to P_1P_2}={p_c^5\over 60\pi m_T^2}g_{T\to P_1P_2}^2, \non \\
 \en
for scalar, vector and tensor mesons, respectively, where $p_c$ is the
c.m. momentum. Numerically, they are given by
\be \label{eq:g}
 && |g^{\rho(770)\to\pi^+\pi^-}|=6.00, \qquad\qquad\quad~
 |g^{K^*(892)\to K^+\pi^-}|=4.59,\non \\
&& |g^{\phi\to K^+K^-}|=4.54\,,   \qquad\qquad\quad\quad~ |g^{\omega\to \pi^+\pi^-}|=0.18\,, \non \\
&& |g^{f_0(980)\to\pi^+\pi^-}|=1.33^{+0.29}_{-0.26}\,{\rm GeV},
\quad |g^{f_0(980)\to K^+K^-}|=3.70\,{\rm GeV},  \\
&& |g^{K_0^*(1430)\to K^+\pi^-}|=3.84\,{\rm GeV}, \quad\quad  |g^{\sigma\to \pi^+\pi^-}|=2.76\,{\rm GeV},  \non \\
&& |g^{f_2(1270)\to\pi^+\pi^-}|=18.56\,{\rm GeV}^{-1}, \quad~ |g^{f_2(1270)\to K^+K^-}|=11.11\,{\rm GeV}^{-1}, \non
\en
where we have used  $\Gamma(f_0(980)\to\pi^+\pi^-)=(34.2^{+13.9+8.8}_{-11.8-2.5})$ MeV \cite{Bellef0}, $m_{\sigma}=563$ MeV and $\Gamma_{\sigma}=350$ MeV obtained in the isobar model fit by the LHCb \cite{Aaij:3pi_2}. Note that the strong coupling constant is determined up to a  strong phase ambiguity, for example, the strong coupling $g^{\sigma\to \pi^+\pi^-}$ has the expression
\be \label{eq:sigmaphi}
g^{\sigma\to \pi^+\pi^-}=|g^{\sigma\to \pi^+\pi^-}|e^{i\phi_\sigma}.
\en
In the below we will use this freedom of the strong phase $\phi_\sigma$ to accommodate a large negative \CP asymmetry through $\pi^+\pi^-\to K^+K^-$ rescattering.

As for the $\rho(1450)$ meson, there is no any experimental information for its decays to $K^+K^-$ and $\pi^+\pi^-$ except for the ratio
\be \label{eq:rho1450R}
R_{\rho(1450)}\equiv {\B(\rho(1450)^0\to K^+K^-)\over \B(\rho(1450)^0\to \pi^+\pi^-)}=0.307\pm0.084\pm0.082\,,
\en
measured by BaBar through the decay $J/\psi\to h^+h^-\pi^0$ \cite{Lees:2017ndy}.
Nevertheless, we can use the measured fractions of $B^-\to \rho(1450)\pi^-\to \pi^+\pi^-\pi^-$ and $K^+K^-\pi^-$ by LHCb and the partial widths of $B^-\to \pi^+\pi^-\pi^-$ and $B^-\to K^+K^-\pi^-$ to extract the strong couplings. Assuming the same $B$--$\rho(1450)$ transition form factors as that of $B$--$\rho(770)$ ones, we obtain
\be \label{eq:rho(1450)coupling}
 g^{\rho(1450)\to K^+K^-}=5.40,\qquad g^{\rho(1450)\to \pi^+\pi^-}=2.31\,.
\en
Contrary to the naive expectation, $\rho(1450)$ couples more strongly to $K^+K^-$ than $\pi^+\pi^-$. This is not consistent with the BaBar's measurement given in Eq. (\ref{eq:rho1450R}).
Since
\be
R_{\rho(1450)}=\left( {g^{\rho(1450)\to K^+K^-}\over g^{\rho(1450)\to \pi^+\pi^-} }\right)^2
\left({ m_{\rho(1450)}^2-4m_K^2 \over m_{\rho(1450)}^2-4m_\pi^2}\right)^{3/2},
\en
it follows that $g^{\rho(1450)\to K^+K^-}/g^{\rho(1450)\to \pi^+\pi^-}\approx 0.85$, in sharp contrast to Eq. (\ref{eq:rho(1450)coupling}).

As we will see in the next section, the decay $B^-\to \rho(1450)^0\pi^-\to \pi^+\pi^-\pi^-$ is well described by the pQCD approach. Hence, the issue has to do with the enormously large coupling of $\rho(1450)$ with $K\ov K$.
Indeed, a recent study in \cite{Wang:2020plx} showed that the pQCD prediction for the branching fraction of $B^+\to\pi^+\rho(1450)^0\to\pi^+K^+K^-$ is about 18 times smaller than experiment.
Note that both BaBar and Belle used to see a broad scalar resonance $f_X(1500)$ in $B\to K^+K^+K^-$, $K^+K^-K_S$ and $K^+K^-\pi^+$ decays at energies around 1.5 GeV. However, the nature of $f_X(1500)$ is not clear as it cannot be identified with the well known scaler meson $f_0(1500)$.
An angular-momentum analysis of the above-mentioned three channels by BaBar \cite{BaBarKKK} showed that the $f_X(1500)$ state is not a single scalar resonance, but instead can be described by the sum of the well-established resonances $f_0(1500)$, $f_0(1710)$ and $f'_2(1525)$. Since $\rho(1450)$ is very board with a width $400\pm60$ MeV \cite{PDG}, a broad vector resonance $\rho_X(1500)$ instead of the scalar one $f_X(1500)$ is an interesting possibility to describe the broad resonance observed at energies $\sim 1.5$ GeV in $B\to KKK$ and $K\ov K\pi$ decays.

\begin{table}
\caption{Branching fractions (in units of $10^{-6}$) and \CP violation of various contributions to $B^\pm\to \pi^\pm K^+ K^-$ decays. The experimental branching fraction of each contribution is inferred from
the measured fit fraction \cite{Aaij:piKK} together with the world average $\B(B^\pm\to \pi^\pm K^+ K^-)=(5.24\pm0.42)\times 10^{-6}$ \cite{HFLAV}, for example, $\B(B^-\to K^*(890)^0K^-\to K^+\pi^-K^-)=(0.39\pm0.05)\times 10^{-6}$. For rescattering contributions, we consider two cases for the $S$-wave $\pi\pi\to K\ov K$ transition amplitudes: Eq. (\ref{eq:trans_1}) for case (i) and Eq. (\ref{eq:trans_2}) for case (ii).
}
\label{tab:KKpi_theory}
\begin{center}
\begin{tabular}{ l c c c c} \hline \hline
 Contribution~~~ & $\B_{\rm expt}$ & ~~~$\B_{\rm theory}$~~~ & $(\A_{C\!P})_{\rm expt}(\%)$~~ & $(\A_{C\!P})_{\rm theory}(\%)$ \\
\hline
 $K^*(890)^0$~~ & ~$0.39\pm0.05$~ & ~~~~$0.23^{+0.04}_{-0.04}$~~~~ & $12.3\pm9.8$ & $-23.7^{+0.2}_{-0.2}$ \\
 $K_0^*(1430)^0$~~ & ~$0.23\pm0.08$~ & $0.71^{+0.13}_{-0.12}$ & $10.4\pm17.3$ & $-19.9^{+0.1}_{-0.1}$ \\
 $\rho(1450)^0$ & ~$1.61\pm0.15$~ & fit & $-10.9\pm5.0$ & $11.4^{+0.3}_{-0.4}$ \\
 $f_2(1270)$ & ~$0.39\pm0.06$~ &$0.05^{+0.01}_{-0.01}$  & $26.7\pm11.3$ & $24.9^{+0.1}_{-0.1}$ \\
 $\phi(1020)$ & ~$0.016\pm0.008$~ & $0.0079^{+0.0019}_{-0.0017}$ & $9.8\pm51.1$ & $0$ \\
 $f_0(980)$ & -- & $0.19^{+0.03}_{-0.03}$ & -- & $15.1^{+0.4}_{-0.5}$ \\
 NR($\pi^\pm K^\mp$) & ~$1.69\pm0.27$~ & $1.68^{+0.43}_{-0.39}$ & $-10.7\pm6.4$ & $-17.8^{+0.1}_{-0.1}$ \\
 NR($K^+K^-$) & -- & $0.14^{+0.05}_{-0.06}$ & -- & $-2.89^{+0.02}_{-0.02}$  \\
 Rescattering & $0.85\pm0.10$ &  (i) $0.75^{+0.21}_{-0.18}$ & $-66.4\pm4.2$ &  fit \\
  &  &  (ii) $0.20^{+0.06}_{-0.05}$ & $-66.4\pm4.2$ &  fit \\
\hline \hline
\end{tabular}
\end{center}
\end{table}

The calculated branching fractions of resonant and nonresonant
contributions to $B^-\to \pi^-K^+K^-$ are summarized in
Table \ref{tab:KKpi_theory}. The theoretical errors arise
from the uncertainties in (i) form factors and the strange quark mass $m_s$,
(ii) the unitarity angle $\gamma$ and (iii) the parameter $\sigma_{_{\rm NR}}$  [see Eq. (\ref{eq:sigma})] which governs the nonresonant matrix elements of scalar densities.

\subsubsection{$\phi(1020)$ production}
The $\phi(1020)$ production proceeds through the $b\to d$ penguin diagram. Its signature is very small due to the smallness of the penguin coefficients $a_{3,5,7,9}$, see Eq. (\ref{eq:FSI}). Indeed,
the branching fraction of  the  quasi-two-body decay $B^-\to\phi\pi^-$
is expected to be very small, of order $4.3\times 10^{-8}$ \cite{CC:Bud}. It is induced mainly from $B^-\to\omega\pi^-$ followed by a small $\omega-\phi$ mixing.
A recent pQCD calculation yields $\B(B^-\to\phi(1020)\pi^+\to K^+K^-\pi^-)=(3.59\pm1.17\pm1.87\pm0.34)\times 10^{-9}$ \cite{Fan:2020gvr}, to be compared with ours $(7.9^{+1.9}_{-1.7})\times 10^{-9}$.

\subsubsection{$K_0^*(1430)$ contribution}
We see from Table \ref{tab:KKpi_theory} that the $K^*_0(1430)^0$ contribution to $B^-\to K^+K^-\pi^-$ is larger than experiment by a factor of 3. Under the narrow width approximation
\be \label{eq:NWA}
\B(B\to RP_3\to P_1P_2P_3)=\B(B\to RP_3)\B(R\to P_1P_2),
\en
the branching fraction \footnote{Since $K_0^*(1430)$ with a width $270\pm80$ MeV is not so narrow, the narrow width approximation is not fully justified and presumably finite-width effects need to be taken into account to extract the branching fraction of $B^-\to K_0^*(1430)^0 K^-$.}
\be
\B(B^-\to K_0^*(1430)^0 K^-)=(0.38\pm0.12\pm0.05)\times 10^{-6}
\en
is obtained by the PDG \cite{PDG}. This mode has been studied in both pQCD and QCDF approaches with the predictions
\be
\B(B^-\to K_0^*(1430)^0 K^-)\times 10^6=1.2^{+0.2+0.1+0.1+0.2}_{-0.1-0.1-0.1-0.2}~(S1),\quad
2.2^{+0.6+0.2+0.4+0.5}_{-0.4-0.2-0.1-0.4}~(S2),
\en
in pQCD \cite{Liu:2010zg} and
\be
\B(B^-\to K_0^*(1430)^0 K^-)\times 10^7=23.71^{+6.67+6.73+2.61}_{-5.60-4.61-3.64}~(S1),\quad
33.70^{+10.33+5.52+3.37}_{-~8.47-4.82-3.94}~(S2),
\en
in QCDF \cite{Li:2015zra}, where S1 and S2 denote two different scenarios for the quark content of the scalar meson. All scalar mesons are made of $q\bar q$ quarks in scenario 1, while in scenario 2 the scalar mesons above 1 GeV are lowest-lying $q\bar q$ scalar states and the light scalar mesons are four-quark states. As discussed in \cite{CCY:SP,Cheng:scalar}, scenario 2 is preferable. It appears that the current theoretical predictions for $\B(B^-\to K_0^*(1430)^0 K^-)$ are too large compared to experiment. This issue needs to be resolved.
It is interesting to notice that the predicted $K^*_0\pi$ rates in $B\to K\pi\pi$ decays are usually {\it smaller} than the results obtained by BaBar and Belle, see Table VI of \cite{Cheng:2013dua}. For example, the calculated branching fraction of $\ov K^{*0}_0(1430)\pi^-$ in $B^-\to K^-\pi^+\pi^-$ is smaller than the BaBar measurement by a factor of two and the Belle result by a factor of three. As discussed in detail in \cite{Cheng:2013dua}, BaBar and Belle have different definitions for the $K_0^*(1430)$ and nonresonant components.

\subsubsection{$f_2(1270)$}
The calculated branching fraction for $f_2(1270)$ is smaller than experiment by a factor of $\sim 7$ in its central value. We have used the form factor $A_0^{Bf_2(1270)}(0)=0.13\pm0.02$ derived from large energy effective theory (see Table II of \cite{Cheng:TP}). Notice that the same form factor leads to a prediction of $\B(B^-\to f_2(1270)\pi^-\to\pi^+\pi^-\pi^-)$ consistent with the experimental value (see Table \ref{tab:3pitheory}). Using the narrow width approximation (\ref{eq:NWA})
and the branching fractions of $f_2(1270)$ \cite{PDG}
\be
\B(f_2(1270)\to K^+K^-)={1\over 2}\times (0.046^{+0.005}_{-0.004}), \quad
\B(f_2(1270)\to \pi^+\pi^-)={2\over 3}\times (0.842^{+0.029}_{-0.009}),
\en
it is straightforward to obtain
\be \label{eq:BFf2pi}
\B(B^-\to f_2(1270)\pi^-)=\cases{ (17.1\pm3.2)\times 10^{-6} & from~$B^-\to f_2(1270)\pi^-\to K^+K^-\pi^-$  \cr (2.4\pm0.5)\times 10^{-6} & from~$B^-\to f_2(1270)\pi^-\to \pi^+\pi^-\pi^-$, }
\en
where the rates of $B^-\to f_2(1270)\pi^-\to K^+K^-\pi^-$ and $B^-\to f_2(1270)\pi^-\to \pi^+\pi^-\pi^-$ are shown in Tables \ref{tab:KKpi_theory} and \ref{tab:3pitheory}, respectively. Evidently, $\B(B^-\to f_2(1270)\pi^-)$ extracted from two different processes differs by a factor of seven! This implies that the $f_2(1270)$ contribution to $B^-\to K^+K^-\pi^-$ is probably largely overestimated experimentally. Indeed, $B^-\to f_2(1270)\pi^-$ is predicted to have the branching fraction of $(2.7^{+1.4}_{-1.2})\times 10^{-6}$ in the QCDF approach \cite{Cheng:TP}. This issue needs to be clarified in the Run II experiment.
(iv) The predicted \CP asymmetry of 25\% in the $f_2(1270)$ component agrees with the measured value,
though the experimental signature for \CP violation is only 2.4$\sigma$. Nevertheless, a large \CP asymmetry is clearly observed in the process of $B^-\to f_2(1270)\pi^-\to \pi^+\pi^-\pi^-$ to be discussed in Sec.III.

\subsubsection{Nonresonant contributions}
Although the nonresonant contribution in the $K^+K^-$ system was not considered by the LHCb, our calculation shown in Table \ref{tab:KKpi_theory} indicates that it is very suppressed relative to the nonresonant one in the $\pi^-K^+$ system. This is contrary to the previous expectation that  the dominant nonresonant contributions for tree-dominated three-body decays arise from the $b\to u$ tree transition rather than from the penguin amplitude process.
We have identified the nonresonant contribution in the $\pi^\pm K^\mp$ system with the matrix element of scalar density $\langle \pi^-K^+|\bar ds|0\rangle^{\rm NR}$. The values of the NR parameters $\alpha_{\rm NR}$, $\sigma_{\rm NR}$ and $\alpha$ in Eqs. (\ref{eq:ADalitz}) and (\ref{eq:sigma}) have been modified in this work.

\subsubsection{\CP violation via rescattering}
From Eqs. (\ref{eq:Ampres}) and (\ref{eq:Swave}), the $S$-wave $\pi^+\pi^-\to K^+K^-$ transition amplitude reads
\be \label{eq:trans_1}
  ie^{i\delta_{\pi\pi}}\sin(\phi/2)(A_{\rm NR}^{\pi^+\pi^-}+A_\sigma).
\en
Recall that the phase $\phi_\sigma$ of the coupling $g^{\sigma\to\pi^+\pi^-}$ is unknown [see Eq. (\ref{eq:sigmaphi})]. By varying $\phi_\sigma$ or the relative phase between $A_\sigma$ and $A_{\rm NR}^{\pi^+\pi^+}$, we find that a large \CP asymmetry of $-66\%$ can be accommodated at $\phi_\sigma\approx 134^\circ$. The branching fraction is $(0.20^{+0.06}_{-0.05})\times 10^{-6}$ as shown in Table \ref{tab:KKpi_theory}.
Since the LHCb analysis of $\pi\pi\leftrightarrow K\ov K$ rescattering is based on the model described in \cite{Bediaga:2015,Bediaga:2013ela},  the $S$-wave transition amplitude in this case is given by
\be \label{eq:trans_2}
  ie^{2i\delta_{\pi\pi}}\sqrt{1-\eta^2}(A_{\rm NR}^{\pi^+\pi^-}+ A_\sigma).
\en
The observed \CP asymmetry is fitted with the same phase $\phi_\sigma=134^\circ$, but the corresponding branching fraction becomes $(0.75^{+0.21}_{-0.18})\times 10^{-6}$. This is consistent with the experimental value of $(0.85\pm0.10)\times 10^{-6}$. Note that the calculated rate for rescattering differs by a factor of $\sim 4$ as the transition amplitude is different by a factor of two to the leading order of $t_{\lambda,\lambda'}$ (see Eqs. (\ref{eq:AmpS}) and (\ref{eq:AmpS1/2})). Nevertheless, we have stressed in passing that one should use Eq. (\ref{eq:FSI}) to describe $\pi\pi\leftrightarrow K\ov K$ final-state rescattering. Therefore, the branching fraction of the rescattering contribution seems to be overestimated by the LHCb by a factor of 4!

\begin{table}[t]
\caption{Direct \CP asymmetries (in \%) and branching fractions of $B^\pm\to \pi^\pm K^+ K^-$ decays with the superscripts denoting ``incl", ``resc" and ``low" for \CP asymmetries measured in full phase space, in the rescattering regions with 1.0 $<m_{K^+K^-}<$ 1.5 GeV and in the low invariant mass region where $m_{K^+K^-}<$ 1.22 GeV, respectively. We consider two cases for the phase of the matrix element $\la \pi^- K^+|\bar ds|0\ra^{\rm NR}$: (i) $\phi_{\pi\!K}=0$ and (ii) $\phi_{\pi\!K}=250^\circ$.
Data are taken from  \cite{LHCb:pippippim} for $\A_{C\!P}^{\rm low}$, \cite{LHCb:2014} for $\A_{C\!P}^{\rm resc}$, \cite{HFLAV} for $\A_{C\!P}^{\rm incl}$ and  $\B(B^-\to \pi^-K^+ K^-)$.
}
\label{tab:KKpiCP}
\begin{ruledtabular}
\begin{tabular}{l  c c c c }
   & $\A_{C\!P}^{\rm incl}$ & $\A_{C\!P}^{\rm resc}$ & $\A_{C\!P}^{\rm low}$~~ & $\B(10^{-6})$ \\
\hline
 Theory with $\phi_{\pi\!K}=0$  & ~$-0.7^{+0.9}_{-0.7}$~ & $13.8^{+1.3}_{-1.2}$ & $15.9^{+1.1}_{-1.0}$ & $4.46^{+0.95}_{-0.85}$ \\
 Theory with $\phi_{\pi\!K}=250^\circ$  & ~$-21.9^{+1.3}_{-1.0}$~ & $-28.6^{+0.3}_{-0.1}$ & $-51.1^{+1.6}_{-1.1}$ & $5.21^{+1.14}_{-1.02}$ \\
 Expt & $-12.3\pm2.1$ & $-32.8\pm4.1$ & $-64.8\pm7.2$ & $5.24\pm0.42$ \\
\end{tabular}
\end{ruledtabular}
\end{table}

\subsubsection{Inclusive and local \CP asymmetries}
The inclusive \CP asymmetry $\A_{C\!P}^{\rm incl}$ in $B^-\to K^+K^-\pi^-$ has been measured at $B$ factories and LHCb with the results: $0.00\pm0.10\pm0.03$ by BaBar \cite{BaBarKpKmpim}, $(-17.0\pm7.3\pm1.7)\%$ by Belle \cite{BelleKpKmpim} and $(-12.3\pm1.7\pm1.2\pm0.7)\%$ by LHCb \cite{LHCb:pippippim}. The world average is $\A_{C\!P}^{\rm incl}=-0.122\pm0.021$ \cite{HFLAV}. Regional \CP asymmetries were also measured by Belle and LHCb. The LHCb measurements read \cite{LHCb:pippippim}
\be
\A_{C\!P}^{\rm low} &=& (-64.8\pm7.0\pm1.3\pm0.7)\%~~{\rm for}~m_{K^+K^-}<1.22\,{\rm GeV}, \non \\
\A_{C\!P}^{\rm resc}&=& (-32.8\pm2.8\pm2.9\pm0.7)\%~~{\rm in}~1.0<m_{K^+K^-}<1.5\,{\rm GeV},
\en
while Belle found \cite{BelleKpKmpim}
\be
\A_{C\!P}^{\rm local}=\cases{ -0.90\pm0.17\pm0.04, & $0.8<m_{K^+K^-}<1.1\,{\rm GeV}$, \cr
-0.16\pm0.10\pm0.01, & $1.1<m_{K^+K^-}<1.5\,{\rm GeV}$, \cr}
\en
and hence a $4.8\sigma$ evidence of a negative \CP asymmetry in the region $m_{K\bar K}<1.1$ GeV.  Note that Belle and LHCb results for local \CP violation are consistent with each other.

In Table \ref{tab:KKpiCP} we show the calculated inclusive and regional \CP asymmetries in the presence of final-state rescattering of $S$-wave $\pi^+\pi^-$ to $K^+K^-$ and compare with experiment. Consider the phase $\phi_{\pi\!K}$ of the matrix element $\la \pi^- K^+|\bar ds|0\ra^{\rm NR}$ defined in Eq. (\ref{eq:KpimeNew2}).  If $\phi_{\pi\!K}=0$ is set to zero, the predicted \CP asymmetries $\A_{C\!P}^{\rm resc}$  and $\A_{C\!P}^{\rm low}$ will be positive, while experimentally they are negative. At first sight, this appears to be a surprise in view of a large and  negative \CP violation coming from rescattering. However, since the branching fraction of  $\pi^+\pi^-\to K^+K^-$ transition is very small, of order $0.2\times 10^{-6}$, its effect can be easily washed out by the presence of various resonances.
Indeed, in our previous work \cite{Cheng:2013dua,Cheng:2016shb} we have considered the case with $\phi_{\pi\!K}=(5/4)\pi$. As shown in Table \ref{tab:KKpiCP}, the agreement between theory and experiment is greatly improved for $\phi_{\pi\!K}\approx 250^\circ$.
It should be stressed that although \CP violation produced by rescattering alone is quite large, of order $-66\%$, the regional \CP asymmetry $\A_{C\!P}^{\rm resc}$ will not be the same as the latter does receive contributions from other resonances.

\section{$B^\pm\to \pi^\pm\pi^+\pi^-$ Decays}

As mentioned in the Introduction, BaBar has carried out the amplitude analysis of $B^-\to \pi^+\pi^-\pi^-$ before \cite{BaBarpipipi}. The nonresonant $S$-wave fraction was measured to be $(34.9\pm4.2^{+8.0}_{-4.5})\%$. In the recent LHCb analysis \cite{Aaij:3pi_1,Aaij:3pi_2}, the $S$-wave component of $B^-\to \pi^+\pi^-\pi^-$ was studied using three different approaches: the isobar model, the $K$-matrix model and a quasi-model-independent (QMI) binned approach. In the isobar model, the $S$-wave amplitude was presented by LHCb as a coherent sum of the $\sigma$ meson contribution and a $\pi\pi\leftrightarrow K\ov K$ rescattering amplitude in the mass range $1.0<m_{\pi^+\pi^-}<1.5$ GeV. The fit fraction of the $S$-wave is about 25\% and predominated by the $\sigma$ resonance (see Table \ref{tab:Data_3pi}). A  large and positive \CP asymmetry of $45\%$ was found in the rescattering amplitude of $B^-\to\pi^+\pi^-\pi^-$, while the corresponding \CP violation in $B^-\to K^+K^-\pi^-$ was of order $-0.66$.

\begin{table}[t]
\caption{Experimental results of the Dalitz plot fit for $B^\pm\to \pi^\pm \pi^+ \pi^-$ decays analyzed in the isobar model \cite{Aaij:3pi_1,Aaij:3pi_2}. }
\label{tab:Data_3pi}
\begin{center}
\begin{tabular}{ l c c r r } \hline \hline
 Contribution~~~ & ~Fit fraction (\%)~~~ & ~~$\A_{C\!P}(\%)$ & ~~~$B^+$ phase ($^\circ$) & ~~$B^-$ phase ($^\circ$)\\
\hline
 $\rho(770)^0$~~ & ~$55.5\pm0.6\pm2.5$~~~ & $0.7\pm1.1\pm1.6$ & --~~~~~~ & --~~~~~~ \\
 $\omega(782)$~~ & ~$0.50\pm0.03\pm0.05$~~~ & $-4.8\pm6.5\pm3.8$ & ~~$-19\pm6\pm1$ & ~~$8\pm6\pm1$ \\
 $f_2(1270)$ & ~$9.0\pm0.3\pm1.5$~~~ & $46.8\pm6.1\pm4.7$ & ~~$5\pm3\pm12$ & ~~$53\pm2\pm12$ \\
 $\rho(1450)^0$ & ~$5.2\pm0.3\pm1.9$~~~ & $-12.9\pm3.3\pm35.9$ & ~~$127\pm4\pm21$ & ~~$154\pm4\pm6$ \\
 $\rho_3(1690)^0$ & ~$0.5\pm0.1\pm0.3$~~~ & $-80.1\pm11.4\pm25.3$ & ~~$-26\pm7\pm14$ & ~~$-47\pm18\pm25$ \\
 S-wave & ~$25.4\pm0.5\pm3.6$~~~ & $14.4\pm1.8\pm2.1$ & --~~~~~~ & --~~~~~~ \\
 ~~Rescattering & ~$1.4\pm0.1\pm0.5$~~~ & $44.7\pm8.6\pm17.3$ & ~~$-35\pm6\pm10$ & ~~$-4\pm4\pm25$ \\
 ~~$\sigma$ & ~$25.2\pm0.5\pm5.0$~~~ & $16.0\pm1.7\pm2.2$ & ~~$115\pm2\pm14$ & ~~$179\pm1\pm95$ \\
\hline \hline
\end{tabular}
\end{center}
\end{table}

Contrary to the decay $B^-\to K^+K^-\pi^-$ where \CP violation is observed only in the rescattering amplitude, a clear \CP asymmetry was seen in the $B^-\to \pi^+\pi^-\pi^-$ decay in the following places: (i) the $S$-wave amplitude at values of $m_{\pi^+\pi^-}$ below the mass of the $\rho(770)$ resonance, see the left panel of Fig. \ref{fig:Fig1}, (ii) the $f_2(1270)$ contribution, see Fig. \ref{fig:Fig1} at values of $m_{\pi^+\pi^-}$ in the $f_2(1270)$ mass region, and (iii) the interference between $S$- and $P$-waves which is clearly visible in Fig. \ref{fig:Fig2} where the data are split according to the sign of $\cos\theta_{\rm hel}$.
In the isobar model, the $S$-wave amplitude is predominated by the $\sigma$ meson. Hence,
a significant \CP violation of 15\%  in $B^-\to\sigma\pi^-$ is implied in this model. The significance of \CP violation in $B^-\to f_2(1270)\pi^-$ was found to be $20\sigma$, $15\sigma$ and $14\sigma$ for the isobar, $K$-matrix and QMI approaches, respectively. Therefore, \CP asymmetry in the $f_2(1270)$ component was firmly established. As for the significance of \CP violation in the interference between $S$- and $P$-waves exceeds $25\sigma$ in all the $S$-wave models.

In contrast to the above-mentioned \CP-violating observables, \CP asymmetry for the dominant quasi-two-body decay mode $B^-\to\rho^0\pi^-$ was found to  be consistent with zero in all three $S$-wave approaches (see Table \ref{tab:CPdata}),  which was already noticed by the LHCb previously in 2014 \cite{LHCb:2014}. However, all the existing theoretical
predictions lead to a negative \CP asymmetry ranging from $-7\%$ to $-45\%$. This is a long-standing puzzle \cite{Cheng:2016shb}.
In this section, we will discuss the observed \CP violation in various modes and address the \CP puzzle with $B^-\to\rho^0\pi^-$.

\begin{table}[t]
\caption{\CP asymmetries in the quasi-two-body decay $B^-\to\rho^0(770)\pi^-$ measured by the LHCb for each $S$-wave approach \cite{Aaij:3pi_1,Aaij:3pi_2}.  }
\label{tab:CPdata}
\begin{ruledtabular}
\begin{tabular}{ c c c c}
  & isobar & $K$-matrix & QMI  \\
\hline
$ \rho(770)^0$ & $0.7\pm1.1\pm0.6\pm1.5$ & $4.2\pm1.5\pm2.6\pm5.8$ & $4.4\pm1.7\pm2.3\pm1.6$ \\
\end{tabular}
\end{ruledtabular}
\end{table}

\subsection{Resonant contributions}
The explicit expression of the factorizable tree-dominated $B^-\to \pi^-(p_1)\pi^+(p_2)\pi^-(p_3)$ decay amplitude can be found in Eq. (2.4) of \cite{Cheng:2013dua}. Amplitudes from various resonances are listed below:

\vskip 0.4cm
\noindent 1. $\rho(770)^0, \rho(1450)^0$
\be \label{eq:rhoamp}
A_{\rho(770,1450)^0} &=& -{1\over \sqrt{2}}{g^{\rho_i\to \pi^+\pi^-} \over s_{23}-m^2_{\rho_i}+im_{\rho_i}\Gamma_{\rho_i}}(s_{12}-s_{13})\Bigg\{ {f_\pi\over 2}\Big[
2m_{\rho_i}A_0^{B\rho_i}(m_\pi^2)  \non \\
&+&  \left(m_B-m_{\rho_i}-{m_B^2-s_{23}\over m_B+m_{\rho_i}}\right)A_2^{B\rho_i}(m_\pi^2)
\Big]\left[a_1 \delta_{pu}+a^p_4+a_{10}^p-(a^p_6+a^p_8) r_\chi^\pi\right]  \\
&+& m_{\rho_i} f_{\rho_i} F_1^{B\pi}(s_{23})\bigg[a_2\delta_{pu}
-a_4^p+\frac{3}{2}(a_7+a_9)+{1\over 2}a_{10}^p\bigg]\Bigg\} + (s_{23}\leftrightarrow s_{12}), \non
\en
with $\rho_i=\rho(770)^0,\rho(1450)^0$.
Since there are two identical $\pi^-$ mesons $\pi^-(p_1)$ and $\pi^-(p_3)$ in this decay, one
should take into account the identical particle effects. As a result, a factor of ${1\over  2}$ should be put in the decay rate.

\vskip 0.4cm
\noindent 2. $\omega(782)$
\be
A_{\omega(782)} &=& -{1\over \sqrt{2}}{g^{\omega\to \pi^+\pi^-} \over s_{23}-m^2_{\omega}+im_{\omega}\Gamma_{\omega}}(s_{12}-s_{13})\Bigg\{ {f_\pi\over 2}\Big[
2m_{\omega}A_0^{B\omega}(m_\pi^2)  \non \\
&+&  \left(m_B-m_{\omega}-{m_B^2-s_{23}\over m_B+m_{\omega}}\right)A_2^{B\omega}(m_\pi^2)\Big]\left[a_1 \delta_{pu}+a^p_4+a_{10}^p-(a^p_6+a^p_8) r_\chi^\pi\right]  \\
&+& m_{\omega}f_{\omega}F_1^{B\pi}(s_{23})\bigg[a_2\delta_{pu}+2(a_3+a_5)
+a_4^p+\frac{1}{2}(a_7+a_9-a_{10}^p)\bigg]\Bigg\}+ (s_{23}\leftrightarrow s_{12}). \non
\en
The strong decay of $\omega(892)$ to $\pi^+\pi^-$ is isospin-violating and it can occur through $\rho$--$\omega$ mixing. In this work we shall use the measured rate of $\omega\to \pi^+\pi^-$ to fix the coupling of $\omega$ with $\pi\pi$.

\vskip 0.4cm
\noindent 3. $f_2(1270)$
\be
A_{f_2(1270)} &=& \sqrt{2}\,{m_{f_2}\over m_B}{f_\pi g^{f_2\to \pi^+\pi^-}\over s_{23}-m^2_{f_2}+im_{f_2}\Gamma_{f_2}} A_0^{Bf_2}(m_\pi^2)\left[{1\over 3}(|\vec{p}_1||\vec{p}_2|)^2-(\vec{p}_1\cdot\vec{p}_2)^2\right]
\non \\
&\times&
\left[a_1 \delta_{pu}+a^p_4+a_{10}^p-(a^p_6+a^p_8) r_\chi^\pi +\beta_2^p \delta_{pu}+\beta_3^p+\beta^p_{\rm 3,EW}\right]+ (s_{23}\leftrightarrow s_{12}),
\en
where
$|\vec{p}_1|$ has the same expression as that in Eq. (\ref{eq:3momentum}), but $|\vec{p}_2|$ and $|\vec{p}_3|$ are replaced by ${1\over 2}\sqrt{s_{23}-4m_\pi^2}$.

\vskip 0.4cm
\noindent 4. $\sigma/f_0(500)$
\be \label{eq:sigmapipi}
A_{\sigma}&=& {g^{\sigma\to \pi^+\pi^-} \over s_{23}-m^2_{\sigma}+im_{\sigma}\Gamma_{\sigma}}\Bigg\{ -f_\pi(m_B^2-s_{23})F_0^{B\sigma^u}(m_\pi^2)\left[a_1 \delta_{pu}+a^p_4+a_{10}^p-(a^p_6+a^p_8) r_\chi^\pi\right] \non \\
&+& {m_\sigma\over m_b-m_d}\, \bar f^d_{\sigma} (m_B^2-m_\pi^2) F_0^{B\pi}(s_{23})(-2a_6^p+a_8^p)\Bigg\}+ (s_{23}\leftrightarrow s_{12}).
\en
In the approach of QCD factorization \cite{Cheng:scalar,CCY:SP}, the decay amplitude of $B^-\to \sigma\pi^-$  has the expression
\be \label{eq:sigmapi}
A(B^- \to \sigma \pi^- ) &=&
 \frac{G_F}{\sqrt{2}}\sum_{p=u,c}\lambda_p
 \Bigg\{ \left[a_1 \delta_{pu}+a^p_4+a_{10}^p-(a^p_6+a^p_8) r_\chi^\pi \right]_{\sigma\pi} X^{(B\sigma,\pi)} \non \\
 &+&
 \left[a_2\delta_{pu} +2(a_3^p+a_5^p)+{1\over 2}(a_7^p+a_9^p)+a_4^p-{1\over 2}a_{10}^p-(a_6^p-{1\over 2}a_8^p)\bar r^\sigma_\chi\right]_{\pi\sigma}\ov X^{(B\pi,\sigma)}\non \\
 &-& f_Bf_\pi\bar f_{\sigma}^u\bigg[\delta_{pu}b_2(\pi\sigma)+ b_3(\pi\sigma)
 + b_{\rm 3,EW}(\pi\sigma) +(\pi\sigma\to \sigma\pi) \bigg] \Bigg\},
\en
where
\be
X^{(B\sigma,\pi)}=-f_\pi (m_B^2-m_\sigma^2)F_0^{B\sigma^u}(m_\pi^2), \qquad
\ov X^{(B\pi,\sigma)}=\bar f_\sigma (m_B^2-m_\pi^2)F_0^{B\pi}(m_\sigma^2),
\en
and $\bar r_\chi^{\sigma}(\mu)=2m_{\sigma}/m_b(\mu)$.
The order of the arguments of the $a_i^p(M_1M_2)$ and
$b_i(M_1M_2)$ coefficients  is dictated by the subscript $M_1M_2$ given in Eq. (\ref{eq:sigmapi}). Note that $a_i^p(\pi\sigma)$ can be numerically very different from $a_i^p(\sigma\pi)$ except for $a_{6,8}^p$. Comparing Eqs. (\ref{eq:sigmapipi}) and (\ref{eq:sigmapi}), we see that the expressions inside $\{\cdots\}$ are identical except that
some terms are missing in Eq. (\ref{eq:sigmapipi}). Those missing terms arise from vertex corrections, hard spectator interactions and penguin contractions.
These subtitles are beyond the simple factorization approach adapted here.

Since the naive amplitude given by Eq. (\ref{eq:sigmapipi}) leads to a negative \CP asymmetry $-0.015$, while experimentally $\acp(\sigma\pi^-)=(16.0\pm2.8)\%$,
we shall follow QCDF to keep those terms missing in the $\sigma$-emission amplitude,
\be \label{eq:sigmapipi_1}
A_{\sigma}&=& {g^{\sigma\to \pi^+\pi^-} \over s_{23}-m^2_{\sigma}+im_{\sigma}\Gamma_{\sigma}}\Bigg\{ -f_\pi(m_B^2-s_{23})F_0^{B\sigma^u}(m_\pi^2)\left[a_1 \delta_{pu}+a^p_4+a_{10}^p-(a^p_6+a^p_8) r_\chi^\pi\right]_{\sigma\pi}  \non \\
&+& \bar f^d_{\sigma} (m_B^2-m_\pi^2) F_0^{B\pi}(s_{23})\left[a_2\delta_{pu} +2(a_3^p+a_5^p)+{1\over 2}(a_7^p+a_9^p)+a_4^p-{1\over 2}a_{10}^p-(a_6^p-{1\over 2}a_8^p)\bar r^\sigma_\chi\right]_{\pi\sigma}\Bigg\} \non \\
&+&  (s_{23}\leftrightarrow s_{12}).
\en
The numerical values of the flavor operators $a_i^p(M_1M_2)$ for $M_1M_2=\sigma\pi$ and $\pi\sigma$ at the scale $\mu=\ov m_b(\ov m_b)$ are exhibited in Appendix B. It is clear that $a_i^p(\pi\sigma)$ and $a_i^p(\sigma\pi)$ can be very different numerically except for $a_{6,8}^p$.

\vskip 0.4cm
\noindent 5. $f_0(980)$
\vskip 0.2cm
It is straightforward to write down the amplitude for the resonance $f_0(980)$ in analog to that of $f_0(500)/\sigma$:
\be \label{eq:f0pipi}
A_{f_0(980)}&=& {g^{f_0\to \pi^+\pi^-} \over s_{23}-m^2_{f_0}+im_{f_0}\Gamma_{f_0}}\Bigg\{ X^{(Bf_0,\pi)}(m_\pi^2)\left[a_1 \delta_{pu}+a^p_4+a_{10}^p-(a^p_6+a^p_8) r_\chi^\pi\right]_{f_0\pi}  \non \\
&+& \ov X^{(B\pi,f_0)}\left[a_2\delta_{pu} +2(a_3^p+a_5^p)+{1\over 2}(a_7^p+a_9^p)+a_4^p-{1\over 2}a_{10}^p-(a_6^p-{1\over 2}a_8^p)\bar r^{f_0}_\chi\right]_{\pi f_0}\Bigg\} \non \\
&+&  (s_{23}\leftrightarrow s_{12}),
\en
with $X^{(Bf_0,\pi)}$ and $\ov X^{(B\pi,f_0)}$ being given by Eq. (\ref{eq:X}).

\subsection{Nonresonant contributions}
Just as the decay $\B^-\to \pi^-K^+K^-$, the nonresonant amplitude in the $\pi^+\pi^-$ system coming from the current-induced process through the $b\to u$ transition reads
\be \label{eq:NRA3pi}
 A_{\rm current-ind}^{\rm NR}=A_{\rm current-ind}^{\rm
  HMChPT}\,e^{-\alpha_{_{\rm NR}} p_B\cdot(p_2+p_3)}\left[a_1 \delta_{pu}+a^p_4+a_{10}^p-(a^p_6+a^p_8) r_\chi^\pi\right],
\en
with
\be
 A_{\rm current-ind}^{\rm HMChPT}
 &=& -\frac{f_\pi}{2}\left[2 m_\pi^2 r+(m_B^2-s_{23}-m_\pi^2) \omega_+
 +(s_{12}-s_{13}) \omega_-\right]+(s_{23}\leftrightarrow  s_{12}).
\en
Besides the current-induced one, an additional nonresonant contribution can also arise from the penguin amplitude
\be
A^{\rm NR}_{\rm penguin}=\la \pi^-|\bar db|B^-\ra\la \pi^+\pi^-|\bar dd|0\ra^{\rm NR}(-2a_6^p+a_8^p)
\en
through the nonresonant matrix element of scalar density $\la \pi^+\pi^-|\bar dd|0\ra^{\rm NR}$.
In our previous work, we have argued that this nonresonant background from the penguin amplitude is suppressed by the smallness of the penguin Wilson coefficients $a_6$ and $a_8$. This is no longer true in view of the very large nonresonant contribution in the $\pi^- K^+$ system of the decay $B^-\to K^+K^-\pi^-$. The nonresonant amplitude
\be
A_{\rm NR}^{\pi^+\pi^-}=A_{\rm current-ind}^{\rm NR}+A^{\rm NR}_{\rm penguin}
\en
is the one we used in Eq. (\ref{eq:Swave}) for describing final-state $\pi^+\pi^-\to K^+K^-$ rescattering.

\subsection{Final-state rescattering}
The rescattering amplitude reads from Eq. (\ref{eq:FSI}) to be
\be \label{eq:Ampres3pi}
A(B^-\to \pi^+\pi^-\pi^-)_{\rm rescattering} &=& e^{i\delta_{\pi\pi}}\Big[\cos(\phi/2)A(B^-\to \pi^+\pi^-\pi^-)_{\rm S-wave}  \non \\
&+& i\sin(\phi/2)A(B^-\to
K^+K^- \pi^-)_{\rm S-wave}\Big],
\en
where the relevant $S$-wave amplitudes $A(B^-\to \pi^+\pi^-\pi^-)_{\rm S-wave}$ and $A(B^-\to K^+K^-\pi^-)_{\rm S-wave}$ are given in  Eq. (\ref{eq:Swave}).

\begin{table}[t]
\caption{Branching fractions (in units of $10^{-6}$) and \CP violation in various contributions to $B^\pm\to \pi^\pm \pi^+ \pi^-$ decays. Experimental results are taken from the isobar model analysis \cite{Aaij:3pi_1,Aaij:3pi_2}.
The experimental branching fraction of each mode is inferred from
the measured fit fraction \cite{Aaij:3pi_1,Aaij:3pi_2}  together with $\B(B^\pm\to \pi^\pm \pi^+ \pi^-)=(15.2\pm1.4)\times 10^{-6}$ \cite{BaBarpipipi}. For rescattering contributions, we consider two cases for the $S$-wave $K\ov K\to \pi\pi$ transition amplitudes: Eq. (\ref{eq:trans_3}) for case (i) and Eq. (\ref{eq:trans_4}) for case (ii).}

\label{tab:3pitheory}
\begin{center}
\begin{tabular}{ l c c c c} \hline \hline
 Contribution~~~ & $\B_{\rm expt}$ & $\B_{\rm theory}$~~~ & $(\A_{C\!P})_{\rm expt}(\%)$~~ & $(\A_{C\!P})_{\rm theory}(\%)$ \\
\hline
 $\rho(770)^0$~~ & ~$8.44\pm0.87$ & $7.67^{+1.62}_{-1.47}$ & $0.7\pm1.9$ & $11.5^{+0.3}_{-0.4}$ \\
 $\omega(782)$~~ & ~$0.076\pm0.011$~~~ &  $0.103^{+0.024}_{-0.021}$ & $-4.8\pm7.5$ & $-14.0^{+0.1}_{-0.1}$ \\
 $f_0(980)$ & -- & ~$0.13^{+0.02}_{-0.02}$~~ & -- & $14.7^{+0.4}_{-0.5}$ \\
 $f_2(1270)$ & ~$1.37\pm0.26$~~ & $1.09^{+0.32}_{-0.28}$ & $46.8\pm7.7$  & $24.9^{+0.1}_{-0.1}$ \\
 $\rho(1450)^0$ & ~$0.79\pm0.11$~~~& fit  & $-12.9\pm36.1$ & $11.2^{+0.3}_{-0.4}$ \\
 $\rho_3(1690)^0$ & ~$0.076\pm0.031$~~~ & -- & $-80.1\pm27.7$ & --\\
 $\sigma(500)$ & ~$3.83\pm0.84$~~~ & $3.15^{+0.52}_{-0.48}$  & $16.0\pm2.8$ & $14.9^{+0.5}_{-0.6}$ \\
 NR($\pi^+\pi^-$) & -- & $2.26^{+0.72}_{-0.61}$ & -- & $48.4^{+11.4}_{-13.8}$ \\
 Rescattering & ~$0.21\pm0.08$~~~ & (i) $0.22^{+0.03}_{-0.03}$ & $44.7\pm19.3$ & $16.3^{+0.8}_{-0.9}$ \\
  &  & (ii) $0.05^{+0.01}_{-0.01}$ & $44.7\pm19.3$ & $16.3^{+0.8}_{-0.9}$ \\
\hline \hline
\end{tabular}
\end{center}
\end{table}

\subsection{Numerical results and discussions}
Using the input parameters summarized in Appendices A and B and the amplitudes given in Sec.III.A, we show the calculated results in Table \ref{tab:3pitheory}.  In the following we shall discuss each contribution in order.

\subsubsection{Nonresonant component}

Although nonresonant contributions were not specified in the LHCb analysis, the theoretical calculations are similar to that of $B^-\to K^+K^-\pi^-$. We find that the nonresonant background denoted by NR$(\pi^+\pi^-)$ in Table \ref{tab:3pitheory} constitutes about 14\% of the $B^-\to \pi^+\pi^-\pi^-$ rate and is dominated by the matrix element of scalar density $\la\pi^+\pi^-|\bar dd|0\ra$. This together the $\sigma$ resonance accounts for 35\% of the total rate. Indeed, the nonresonant fraction was found to be 35\% in the earlier BaBar measurement \cite{BaBarpipipi}. As discussed in Sec.II.D.5, a large and negative \CP asymmetry in the rescattering amplitude of $B^-\to\pi^- K^+K^-$ cannot be accommodated unless the amplitude $A_\sigma$ interferes with $A_{\rm NR}^{\pi^+\pi^-}$.

\subsubsection{$\omega(782)$}
Since the $\omega(782)$ is very narrow in its width, the factorization relation for three-body decay under the narrow width approximation is expected to be valid
\be \label{eq:fact}
 \B(B^-\to \omega\pi^-\to \pi^+\pi^-\pi^-)=\B(B^-\to \omega\pi^-)\B(\omega\to \pi^+\pi^-).
\en
Using the world average $\B(B^-\to \omega\pi^-)=(6.9\pm0.5)\times 10^{-6}$ \cite{PDG} and the branching fraction $\B(\omega\to \pi^+\pi^-)=(1.53\pm0.06)\%$ \cite{PDG}, it is expected that $\B(B^-\to \omega\pi^-\to \pi^+\pi^-\pi^-)=(0.106\pm0.009)\times 10^{-6}$. This is consistent with both theory and the LHCb measurement.

\subsubsection{$f_2(1270)$}

The calculated branching fraction and \CP asymmetry of 25\% for the process $B^-\to f_2(1270)\pi^-\to\pi^+\pi^-\pi^-$  are in accordance with experiment. Recall that the previous measurement by BaBar yields $\acp(B^-\to f_2(1270)\pi^-)=0.41\pm0.25$ \cite{BaBarpipipi}. \CP asymmetry of $(46.8\pm7.7)\%$ in the $f_2(1270)$ contribution was finally firmly established by the LHCb. We have shown in Eq. (\ref{eq:BFf2pi}) two very different results of $\B(B^-\to f_2(1270)\pi^-)$ extracted from two different processes $B^-\to f_2(1270)\pi^-\to K^+K^-\pi^-$ and $B^-\to f_2(1270)\pi^-\to \pi^+\pi^-\pi^-$. From the latter process, BaBar's measurement yields $\B(B^-\to f_2(1270)\pi^-)=(1.60^{+0.67+0.02}_{-0.44-0.06})\times 10^{-6}$ \cite{PDG}.
This is consistent with the result of  $\B(B^-\to f_2(1270)\pi^-)=(2.4\pm0.5)\times 10^{-6}$
inferred from the LHCb [cf Eq. (\ref{eq:BFf2pi})].

\subsubsection{$\rho(1450)$}
By considering the $P$-wave time-like electromagnetic form factor $F_\pi$ for the charged pions $\pi^+\pi^-$ in the region of $\rho(1450)$ extracted from the available experimental data, the authors of
\cite{Li:rho(1450)} have studied the decay $B^-\to\rho(1450)^0\pi^-\to \pi^+\pi^-\pi^-$ within the pQCD approach. The result $\B(B^-\to\rho(1450)^0\pi^-\to \pi^+\pi^-\pi^-)=(8.15^{+1.46}_{-1.32})\times 10^{-7}$  agrees well with the measured value of $(7.9\pm1.1)\times 10^{-7}$. However, when this approach is generalized to the $P$-wave time-like form factor $F_K$ for the charged kaons $K^+K^-$, it appears that the calculated rate for $B^-\to\rho(1450)^0\pi^-\to K^+K^-\pi^-$ is too small compared to experiment \cite{Wang:2020plx}. This issue with $\rho(1450)\to K^+K^-$ needs to be resolved  in the future.

\subsubsection{$\sigma/f_0(500)$}
Using $m_\sigma=563$ MeV, $\Gamma_\sigma= 350$ MeV, the decay constants and form factors given in Appendix A, the decay amplitude presented in Eq. (\ref{eq:sigmapipi_1}) and the flavor operators $a_i^p(M_1M_2)$ for $M_1M_2=\sigma\pi$ and $\pi\sigma$  shown in Table \ref{tab:aiSP}, the resulant branching fraction $\B(B^-\to\sigma\pi^-\to\pi^+\pi^-\pi^-)=(3.15^{+0.52}_{-0.48})\times 10^{-6}$ and the \CP asymmetry $\acp(\sigma\pi^-)=(14.9^{+0.5}_{-0.6})\%$ are in good agreement with experiment (cf. Table \ref{tab:3pitheory}).  Since $\sigma$ is very broad, its finite width effect which has been considered in \cite{Qi:2018lxy} could be quite important .

\subsubsection{\CP violation via rescattering}
The $S$-wave $K^+K^-\to \pi^+\pi^-$ transition amplitude reads from Eqs. (\ref{eq:Ampres3pi}) and (\ref{eq:Swave}) to be
\be \label{eq:trans_3}
  ie^{i\delta_{\pi\pi}}\sin(\phi/2)(A_{\rm NR}^{K^+K^-}+A_{f_0(980)}^{K^+K^-}).
\en
Since both nonresonant contribution in the $K^+K^-$ system and the $f_0(980)$ contribution to $B^-\to K^+K^-\pi^-$ have not been studied by the LHCb yet, we have to rely on the theoretical evaluation of these two amplitudes. The LHCb measurement of the rescattering contribution to $B^-\to \pi^+\pi^-\pi^-$ corresponds to the following transition amplitude
\be \label{eq:trans_4}
  ie^{2i\delta_{\pi\pi}}\sqrt{1-\eta^2}(A_{\rm NR}^{K^+K^-}+A_{f_0(980)}^{K^+K^-}).
\en
Here we shall adapt a strategy different from that in the decay $B^-\to \pi^+\pi^-\pi^-$. We first vary the phase of the $f_0(980)K^+K^-$ coupling to fit the ``measured" branching fraction and then figure out the \CP asymmetry induced by rescattering. It turns out at $\phi_{f_0(980)}\approx 20^\circ$, the phase of $g^{f_0(980)\to K^+K^-}$, the $K^+K^-\to \pi^+\pi^-$ transition amplitude (\ref{eq:trans_4}) yields $\B({\rm rescattering})=(0.22\pm0.03)\times 10^{-6}$ and a \CP asymmetry of $(16.3^{+0.8}_{-0.9})\%$ (see Table \ref{tab:3pitheory}). For the transition amplitude of Eq. (\ref{eq:trans_3}), the branching fraction becomes smaller by a factor of 4, namely, $(0.05\pm0.01)\times 10^{-6}$.
Therefore, the branching fraction of the rescattering contribution seems to be overestimated experimentally by a factor of $\sim 4$!

\subsubsection{Inclusive and local \CP asymmetries}

In Table \ref{tab:pipipiCP} we show  inclusive and regional \CP asymmetries in $B^\pm\to \pi^\pm\pi^+\pi^-$ decays.
The calculated $\A_{C\!P}^{\rm incl}$ and $\A_{C\!P}^{\rm resc}$ are too large compared to experiment. For a consideration of $\rho$--$\omega$ mixing effect on local \CP violation, see \cite{XHGuo}.

\subsubsection{\CP asymmetry induced by interference}
\begin{figure}[t]
\begin{center}
\includegraphics[width=0.35\textwidth]{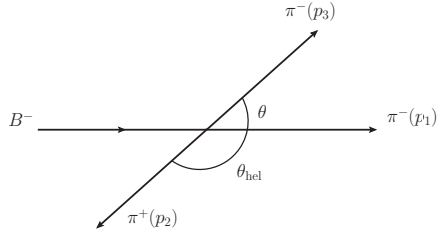}
\vspace{0.1cm}
\caption{The angle $\theta$ between the momenta of the two $\pi^-$ pions measured in the rest frame of the dipion system in the decay $B^-\to \pi^-(p_1)\pi^+(p_2)\pi^-(p_3)$. It is related to the helicity angle $\theta_{\rm hel}$ defined by the LHCb through the relation $\theta+\theta_{\rm hel}=\pi$.} \label{fig:theta}
\end{center}
\end{figure}

Before proceeding to discuss the \CP asymmetry induced by interference, we follow \cite{Bediaga:2015} to define the quantity $\theta$ being the angle between the pions with the same-sign charge. For example, in $B^-\to \pi^-\pi^+\pi^-$ decay, it is the angle between the momenta of the two $\pi^-$ pions measured in the rest frame of the dipion system (i.e. the resonance). This angle is related to the helicity angle $\theta_{\rm hel}$ defined by the LHCb \cite{Aaij:3pi_2} through the relation $\theta_{\rm hel}+\theta=\pi$ (see Fig. \ref{fig:theta}). Hence, $\cos\theta_{\rm hel}=-\cos\theta$.

\begin{table}[t]
\caption{Same as Table \ref{tab:KKpiCP} except for $B^\pm\to \pi^\pm \pi^+ \pi^-$ decays.
}
\vskip 0.2cm
\label{tab:pipipiCP}
\begin{ruledtabular}
\begin{tabular}{l  c c c c }
   & $\A_{C\!P}^{\rm incl}$ & $\A_{C\!P}^{\rm resc}$ & $\A_{C\!P}^{\rm low}$~~ & $\B(10^{-6})$ \\
\hline
 Theory   & ~$28.2^{+0.3}_{-0.5}$~ & $42.4^{+0.3}_{-0.8}$ & $45.5^{+1.9}_{-2.4}$ & $20.4^{+4.5}_{-3.9}$ \\
 Expt & $5.8\pm2.4$ & $17.2\pm2.7$ & $58.4\pm9.7$ & $15.2\pm1.4$ \\
\end{tabular}
\end{ruledtabular}
\end{table}

Consider the decay $B^-\to \pi^-(p_1)\pi^+(p_2)\pi^-(p_3)$ and define $s_{23}=(p_2+p_3)^2=m^2_{\pi^+\pi^- \rm ~low}$. The angular distribution of the vector resonance is governed by the term $s_{12}-s_{13}$ (see, for example, Eq. (\ref{eq:rhoamp})).
From Eq. (\ref{eq:angle}) we have
\be \label{eq:s&theta}
s_{12}-s_{13}=-4\vec{p}_1\cdot\vec{p}_2=-4|\vec{p}_1||\vec{p}_2|\cos\theta_{\rm hel}=4\vec{p}_1\cdot\vec{p}_3=4|\vec{p}_1||\vec{p}_3|\cos\theta
\en
in the rest frame of $\pi^+(p_2)$ and $\pi^-(p_3)$. As noticed in passing, $|\vec{p}_1|$ has the same expression as that in Eq. (\ref{eq:3momentum}), but $|\vec{p}_2|$ and $|\vec{p}_3|$ are replaced by ${1\over 2}\sqrt{s_{23}-4m_\pi^2}$.
Furthermore, it follows from Eq. (\ref{eq:s&theta}) that $\cos\theta$ can be expressed  as a function of $s_{12}$ and $s_{23}$
\be
\cos\theta=a(s_{23})s_{12}+b(s_{23}),
\en
with \cite{Bediaga:2015}
\be
a(s) &=& {1\over (s-4m_\pi^2)^{1/2}\left({(m_B^2-m_\pi^2-s)^2\over 4s} - m_\pi^2\right)^{1/2}}, \non \\
b(s) &=& -{m_B^2+3m_\pi^2-s\over 2(s-4m_\pi^2)^{1/2}\left({(m_B^2-m_\pi^2-s)^2\over 4s} - m_\pi^2\right)^{1/2}}.
\en

\begin{figure}[t]
\begin{center}
\includegraphics[width=0.35\textwidth]{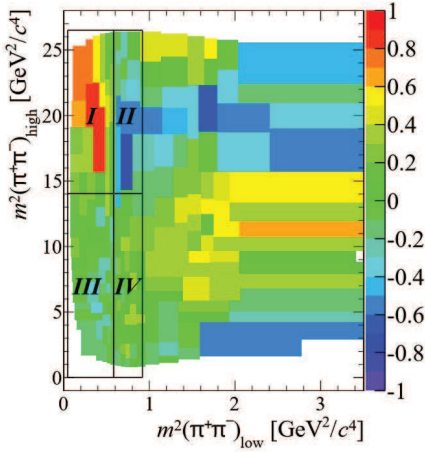}
\vspace{0.1cm}
\caption{The low $\pi^+\pi^-$ invariant mass region of the $B^+\to \pi^+\pi^+\pi^-$ Dalitz plot of \CP asymmetries divided into four zones. This plot is taken from \cite{Reis}.} \label{fig:Dalitz}
\end{center}
\end{figure}

For \CP violation induced by the interference between different resonances, let us consider the low $\pi^+\pi^-$ invariant mass region of the Dalitz plot which is divided into four zones
as shown in Fig. \ref{fig:Dalitz}. The vertical line dividing zones I and III from zones II and IV is at the $\rho(770)$ mass, while the horizontal line separating zones I and II from zones III and IV is at the position where $\cos\theta=0$, corresponding to $s_{12}=-b/a$. The cosine of the angle $\theta$ varies from $-1$ to 0 in zones III and IV, corresponding to $(s_{12})_{\rm min}=-(1+b)/a$ and $s_{12}=-b/a$, respectively. Likewise, The cosine of the angle $\theta$ varies from $0$ to 1 in zones I and II, corresponding to $s_{12}=-b/a$ and
$(s_{12})_{\rm max}=(1-b)/a$, respectively. Hence,
\be
{\rm I, II}: && \int_0^1\cos\theta\, d\!\cos\theta=\int^{(s_{12})_{\rm max}}_{-b/a}(as_{12}+b)\,ds_{12}={1\over 2},  \non \\
{\rm III, IV}: && \int_{-1}^0\cos\theta\, d\!\cos\theta=\int^{-b/a}_{(s_{12})_{\rm min}}(as_{12}+b)\,ds_{12}=-{1\over 2}.
\en
In short, zones I and II are delimited by $\cos\theta>0$ or $\cos\theta_{\rm hel}<0$, while
zones III and IV are delimited by $\cos\theta<0$ or $\cos\theta_{\rm hel}>0$.

The difference in the number of $B^-$ and $B^+$ events measured in  the low-$m_{\rm low}$ region for (a) $\cos\theta<0$ (or $\cos\theta_{\rm hel}>0$) and (b) $\cos\theta>0$ (or $\cos\theta_{\rm hel}<0$) is depicted in Fig. \ref{fig:Fig2}. In Fig. \ref{fig:Fig2}(a) we see that $\A_{C\!P}$ which is proportional to $N_{B^-}-N_{B^+}$ is negative below the $\rho(770)$ mass (zone III) and positive above it (zone IV) with a zero at $m_{\rm low}=m_\rho$, while in Fig. \ref{fig:Fig2}(b) $\A_{C\!P}$ is positive below the $\rho(770)$ mass (zone I) and negative above it (zone II).
The sum of \CP asymmetries of $\cos\theta>0$ and $\cos\theta<0$ gives rise to the \CP violation shown in the left panel of Fig. \ref{fig:Fig1}. It is clear that \CP asymmetry at
$m_{\rm low}$ below the $\rho$ mass is of order 20\%, which is the sum of zone I and zone III. From Fig. \ref{fig:Dalitz} it is evident that the local \CP asymmetry is largest in zone I. Indeed, LHCb has measured
$\A_{C\!P}^{\rm low}(\pi^+\pi^-\pi^-)$ to be $0.584\pm0.082\pm0.027\pm0.007$
in the region specified by $m^2_{\pi^-\pi^- \rm ~low}<0.4$ GeV$^2$ and $m^2_{\pi^+\pi^- \rm ~high}>15$ GeV$^2$ \cite{LHCb:pippippim}.

\begin{figure}[t]
\centering
 {
      \includegraphics[scale=0.65]{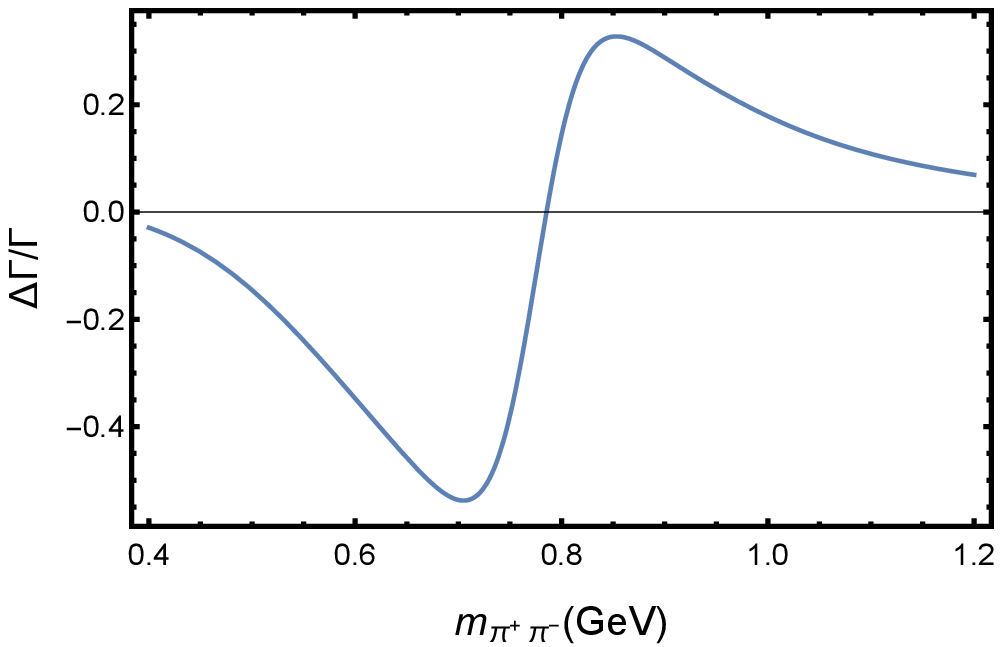}
}{
     \includegraphics[scale=0.65]{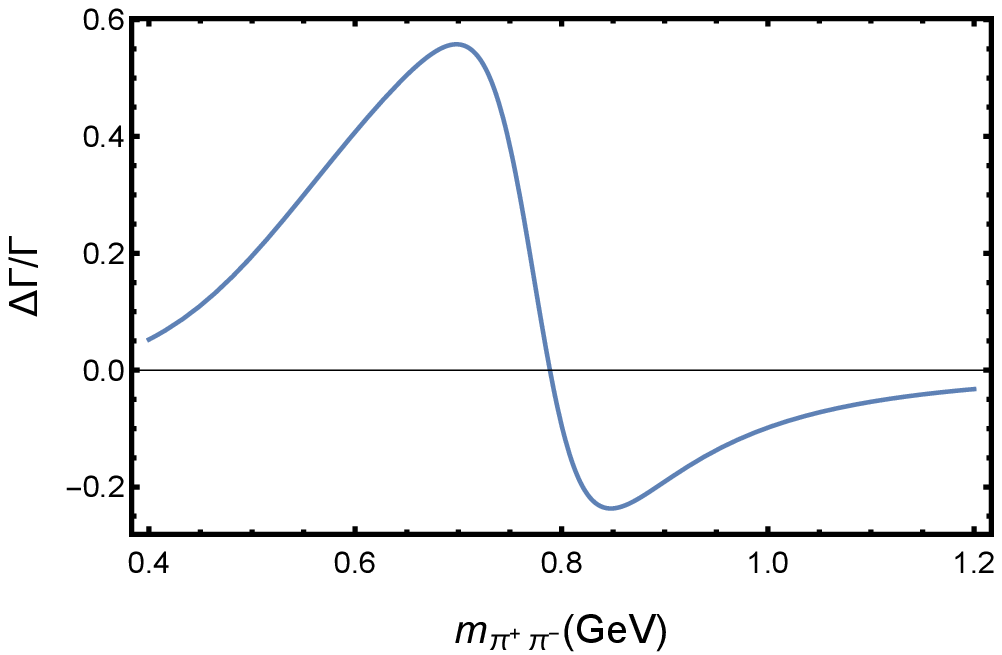}
}
\vskip 0.2cm
\centering{(a) \hskip 5.6cm (b)}
\vskip 0.5cm
 {
      \includegraphics[scale=0.65]{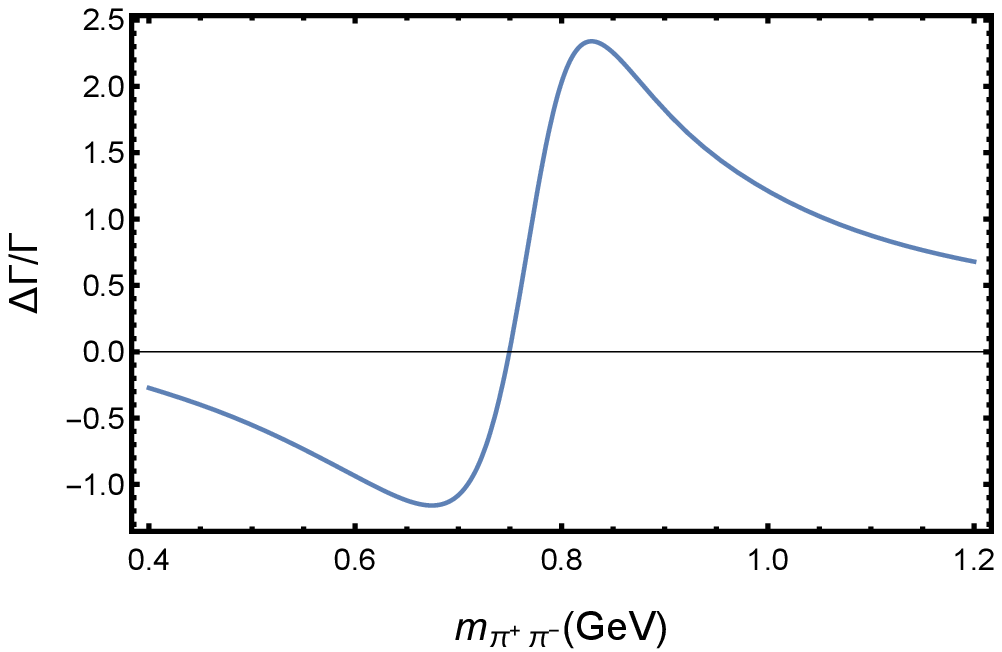}
}{
     \includegraphics[scale=0.65]{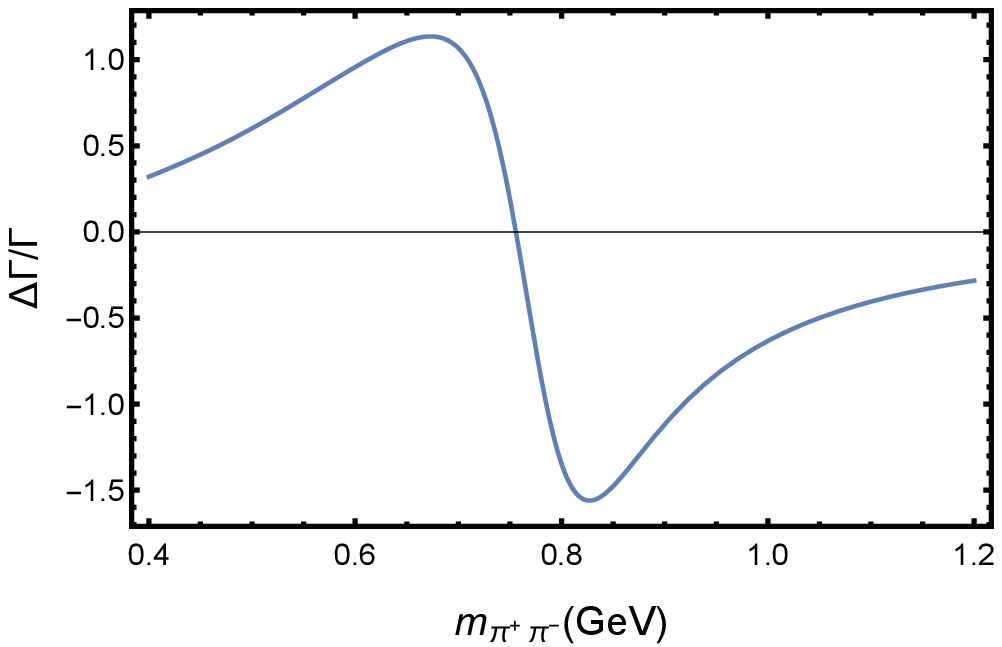}
}
\vskip 0.2cm
\centering{(c) \hskip 5.6cm (d)}
\caption{The rate asymmetry $\Delta\Gamma$ in units of $\Gamma=1/\tau(B^\pm)$ for $B^\pm\to \pi^\pm\pi^+\pi^-$ in the low-$m_{\rm low}$ region induced by the interference between $\rho(770)$ and the $\sigma$ meson for (a) $\cos\theta<0$ or  $\cos\theta_{\rm hel}>0$ and (b) $\cos\theta>0$ or  $\cos\theta_{\rm hel}<0$. The interference between $\rho(770)$ and the nonresonant amplitude is added  to (a) and (b) and shown in (c) and (d), respectively. }
\label{fig:angular_theory}
\end{figure}

In \cite{Bediaga:2015} the \CP asymmetry of the $B^-\to\pi^+\pi^-\pi^-$ decay in the low-mass region with $s_{23}<1$ GeV$^2$ shown in Fig. \ref{fig:Fig2} is described by the interference between the $\rho$ and the nonresonant amplitude  and the interference between the $\rho(770)$ and $f_0(980)$ mesons.
Writing
\be
A_\pm\equiv A^\rho_\pm+A^{\rm NR}_\pm=c_\pm^\rho F_\rho^{\rm BW}\cos\theta+ c_\pm^{\rm NR},
\en
for the $B^+$ and $B^-$ decays, where $F_\rho^{\rm BW}$ is the Breit-Wigner propagator of the $\rho(770)$
\be
  F_\rho^{\rm BW}(s_{23})={1\over s_{23}-m_\rho^2+ im_\rho\Gamma_\rho},
\en
it follows that \CP asymmetry has the expression
\be \label{eq:modelCP}
\A_{C\!P} &\propto& (|c_-^\rho|^2-|c_+^\rho|^2)|F_\rho^{\rm BW}(s_{23})|^2 \cos\theta^2+  (|c_-^{\rm NR}|^2-|c_+^{\rm NR}|^2)  \non \\
&+& 2\,{\rm Re}(c_-^{*\rho}c_-^{\rm NR}-c_+^{*\rho}c_+^{\rm NR})|F_\rho^{\rm BW}(s_{23})|^2(s_{23}-m_\rho^2)\cos\theta  \\
&+& 2\,{\rm Im}(c_-^{*\rho}c_-^{\rm NR}-c_+^{*\rho}c_+^{\rm NR})|F_\rho^{\rm BW}(s_{23})|^2 m_\rho\Gamma_\rho \cos\theta.  \non
\en
The terms  $(s_{23}-m_\rho^2)\cos\theta$ and $m_\rho\Gamma_\rho$ arise from the imaginary and real parts, respectively, of the Breit-Wigner propagator $F_\rho^{\rm BW}$. It was argued in \cite{Bediaga:2015} that the first two terms violate the {\it CPT} constraint locally and will be set to zero. Assuming $c_\pm^\rho$ and $c_\pm^{\rm NR}$ are complex constants, the parameters ${\rm Re}(c_-^{*\rho}c_-^{\rm NR}-c_+^{*\rho}c_+^{\rm NR})$ and ${\rm Im}(c_-^{*\rho}c_-^{\rm NR}-c_+^{*\rho}c_+^{\rm NR})$ were obtained in \cite{Bediaga:2015} by fitting them to the data. The observed interference pattern in the $\rho$ region is mainly described by the $(s_{23}-m_\rho^2)\cos\theta$ term.

Instead of fitting the unknown parameters to the data, we would like to predict the interference pattern in our approach. Since the fit fraction of the broad scalar meson $\sigma$ is about 25\% in the isobar model, it is natural to consider the interference between the $\rho(770)$ and $\sigma(500)$ mesons (or the broad $S$-wave in the other models)
\be
\Gamma^{\rho-\sigma}(s_{23}) &=& {1\over (2\pi)^3 32m_B^3}\,{G_F^2\over 2}\,{1\over 2}\int_{(s_{12})_{\rm min}}^{-b/a}2[{\rm Re}(A_\rho){\rm Re}(A_\sigma)+{\rm Im}(A_\rho){\rm Im}(A_\sigma)]\,ds_{12}~~~{\rm for}~\cos\theta<0,   \non \\
\Gamma^{\rho-\sigma}(s_{23}) &=& {1\over (2\pi)^3 32m_B^3}\,{G_F^2\over 2}\,{1\over 2}\int^{(s_{12})_{\rm max}}_{-b/a}2[{\rm Re}(A_\rho){\rm Re}(A_\sigma)+{\rm Im}(A_\rho){\rm Im}(A_\sigma)]\,ds_{12}~~{\rm for}~\cos\theta>0, \non \\
\en
where the identical particle effect has been taken care of by the factor of 1/2, and the amplitudes
$A_{\rho(770)}$ and $A_\sigma$ are given by Eqs. (\ref{eq:rhoamp}) and (\ref{eq:sigmapipi_1}), respectively. The rate asymmetry $\Delta \Gamma^{\rho-\sigma}\equiv \Gamma_{B^-\to \pi^-\pi^+\pi^-}-\Gamma_{B^+\to\pi^+\pi^+\pi^-}$ due to the $\rho(770)$ and $\sigma$ interference is shown in Figs. \ref{fig:angular_theory}(a) and \ref{fig:angular_theory}(b)
for $\cos\theta<0$ and $\cos\theta>0$, respectively. It is evident that the sign of \CP asymmetry is flipped below and above the $\rho(770)$ peak and that the interference term is proportional to $\cos\theta$. Our calculation indicates that \CP asymmetry is positive in zones I and IV, negative in zones II and III, in agreement with the data (see Fig. \ref{fig:Fig2}). The interference between $\rho$ and the nonresonant amplitude exhibits a similar feature. This interference effect is included
in Figs. \ref{fig:angular_theory}(c) and \ref{fig:angular_theory}(d) with the rate asymmetry
$\Delta \Gamma^{\rho-\sigma,\rho-{\rm NR}}$. Note that \CP violation no longer vanishes exactly at $s_{23}=m_\rho^2$ due to the contributions from the imaginary part of $F_\rho^{\rm BW}$.
In short, the rate asymmetry depicted in Fig. \ref{fig:Fig2} is the first observation of \CP violation mediated by interference between resonances with significance exceeding 25$\sigma$, though it vanishes in the $\rho(770)$ region when integrating over the angle.

\subsubsection{\CP violation in $B^-\to \rho^0\pi^-$}
As noticed in passing, \CP asymmetry for the quasi-two-body decay $B^-\to\rho^0\pi^-$ was found by LHCb to  be consistent with zero in all three $S$-wave approaches (cf. Table \ref{tab:CPdata}).  \footnote{There was a measurement of $\acp(\rho^0\pi^-)$ by BaBar with the result $0.18\pm0.07^{+0.05}_{-0.15}$ from a Dalitz plot analysis of $B^-\to \pi^+\pi^-\pi^-$ \cite{BaBarpipipi}.}
Indeed, if this quasi-two-body \CP asymmetry is nonzero, it will destroy the interference pattern observed in Fig. \ref{fig:Fig2}, see the first term in Eq. (\ref{eq:modelCP}). However, the existing theoretical
predictions based on QCD factorization (QCDF) \cite{CC:Bud,Sun:2014tfa}, perturbative QCD (pQCD) \cite{LiYa:2016}, soft-collinear effective theory (SCET) \cite{Wang:2008rk}, topological diagram approach (TDA) \cite{Cheng:TDA} and factorization-assisted topological-amplitude (FAT) approach \cite{Zhou:2016jkv} all lead to a negative \CP asymmetry for $B^-\to \rho^0\pi^-$, ranging from $-7\%$ to $-45\%$ (see Table \ref{tab:rhopiCP}).

\begin{table}[t]
\caption{Theoretical predictions of \CP violation (in \%) for the $B^-\to \rho^0\pi^-$ decay in various approaches.}
\label{tab:rhopiCP}
\footnotesize{
\begin{ruledtabular}
\begin{tabular}{ c c c c c c}
 QCDF \cite{CC:Bud} & QCDF \cite{Sun:2014tfa} & pQCD \cite{LiYa:2016} & SCET \cite{Wang:2008rk} & TDA \cite{Cheng:TDA} & FAT \cite{Zhou:2016jkv} \\
\hline
$-9.8^{+3.4+11.4}_{-2.6-10.2}$~~ & ~~$-6.7^{+0.2+3.2}_{-0.2-3.7}$~~ & ~~$-27.5^{+2.3+0.9}_{-3.1-1.0}\pm1.4\pm0.9$~~  & ~~$-19.2^{+15.5+1.7}_{-13.4-1.9}$~~ & ~~$-23.9\pm 8.4$~~ & $-45\pm4$\\
\end{tabular}
\end{ruledtabular}}
\end{table}

It has been argued in \cite{Bediaga:2016} that in $B\to PV$ decays with $m_V<1$ GeV, for example, $V=\rho(770)$ or $K^*(892)$, \CP asymmetry induced from a short-distance mechanism is suppressed by the $C\!PT$ constraint. Under the the `2+1' approximation that the resonances produced in heavy meson decays do not interact with the third particle, there do not exist other states which can be connected to $\pi\pi$ or $\pi K$ through final-state interactions. Hence, the absence of final-state interactions implies the impossibility to observe \CP asymmetry in those processes. However,
if we take this argument seriously to explain the approximately vanishing \CP asymmetry in $B^+\to \rho^0\pi^+$, it will be at odd with the \CP violation seen in other $PV$ modeds. For example, \CP violation in the decay $B^0\to K^{*+}\pi^-$ with  $\A_{C\!P}=-0.308\pm0.062$ was clearly observed by the LHCb \cite{LHCb:Kstpi}. Therefore, it appears that the smallness of $\A_{C\!P}(B^+\to\rho^0 \pi^+)$ has nothing to do with the $C\!PT$ constraint.

As elucidated in \cite{Cheng:2020hyj}, the nearly vanishing \CP violation in $B^-\to\rho^0\pi^-$ is understandable in the QCD factorization approach.  There are two kinds of $1/m_b$ corrections in QCDF: penguin annihilation to the penguin amplitude and hard spectator interactions to the flavor operator $a_2$. Power corrections in QCDF often involve endpoint divergences which are parameterized in terms of the parameters $\rho_A$, $\phi_A$ for penguin annihilation and $\rho_H$, $\phi_H$ for hard spectator interactions (see Eq. (\ref{eq:XA}) in Appendix B). In the heavy quark limit,  $\acp(\rho^0\pi^-)$ is of order 6.3\%. Power corrections induced from hard spectator interactions will push it up further, say
$\acp(\rho^0\pi^-)\sim 15\%$, whereas penguin annihilation will pull it to the opposite direction (see Table III of \cite{Cheng:2020hyj}). Owing to the destructive contributions from these two different $1/m_b$ power corrections, a nearly vanishing  $\acp(\rho^0\pi^-)$ can be {\it accommodated} in QCDF. For example, $\B(\rho^0\pi^-)\approx 8.4\times 10^{-6}$ and $\acp(\rho^0\pi^-)\approx (-0.7^{+5.4}_{-4.5})\%$ are obtained with $(\rho_H, \rho_A^i,\rho_A^f)=(3.15,3.08,0.83)$ and $(\phi_H, \phi_A^i,\phi_A^f)=(-113^\circ,-145^\circ,-36^\circ)$  \cite{Cheng:2020hyj}, while experimentally $\B(\rho^0\pi^-)=(8.3^{+1.2}_{-1.3})\times 10^{-6}$ \cite{HFLAV} and $\acp(\rho^0\pi^-)= (0.7\pm1.9)\%$ in the isobar model.

\subsubsection{\CP violation at high $m_{\rm high}$}
An inspection of Fig. \ref{fig:Fig1} for \CP asymmetries measured in the high invariant-masss region, the peak in the high-$m_{\rm high}$ region could be ascribed to the $\chi_{c0}(1P)$ resonance with a mass $3414.71\pm0.30$ MeV and a width $10.8\pm0.6$ MeV. As stressed in \cite{Bediaga:2020ztp}, although LHCb has not yet found the contribution from the $B^-\to \pi^-\chi_{c0}$ amplitude in $B^-\to \pi^+\pi^-\pi^-$ decay, the Mirandizing distribution for Run I data has already shown a clear and huge \CP asymmetry around the $\chi_{c0}$ invariant mass. We also see from Fig. \ref{fig:Fig1} that \CP asymmetry in the high-$m_{\rm high}$ region changes sign at around 4 GeV, near the $D\ov D$ threshold. In analog to the $\pi\pi\leftrightarrow K\ov K$ rescattering in the low mass region, final-state rescattering $D\ov D\to P\ov P$ could provide the strong phases necessary for \CP violation in the high-$m_{\rm high}$ region \cite{Bediaga:2020ztp,Mannel:2020}. However, we will not address this issue in this work.

\section{Conclusions}
We have presented in this work a study of charmless three-body decays of $B$ mesons
$B^-\to K^+K^-\pi^-$ and $B^-\to\pi^+\pi^-\pi^-$
based on the factorization approach. Our main results are:

\begin{itemize}

\item There are two distinct sources of nonresonant contributions: one arises from
from the $b\to u$ tree transition and the other from the nonresonant matrix element of scalar densities $\la M_1M_2|\bar q_1 q_2|0\ra^{\rm NR}$. It turns out that even for tree-dominated three-body decays $B\to \pi\pi\pi$ and $K\ov K\pi$, nonresonant contributions are dominated by the penguin mechanism rather than by the $b\to u$ tree process, as  implied by the large nonresonant component observed in the $\pi^- K^+$ system which accounts for one third of the $B^-\to K^+K^-\pi^-$ rate. We have identified the nonresonant contribution to the $\pi^- K^+$ system  with the matrix element $\langle \pi^-K^+|\bar ds|0\rangle^{\rm NR}$.

\item  The calculated branching fraction of  $B^-\to f_2(1270)\pi^-\to K^+K^-\pi^-$ is smaller than experiment by a factor of $\sim 7$ in its central value.  Nevertheless, the same form factor for $B\to f_2(1270)$ transition leads to a prediction of $\B(B^-\to f_2(1270)\pi^-\to\pi^+\pi^-\pi^-)$ in agreement with the experimental value.
    Branching fractions of  $B^-\to f_2(1270)\pi^-$ extracted from the measured rates of
    $B^-\to f_2(1270)\pi^-\to K^+K^-\pi^-$ and $B^-\to f_2(1270)\pi^-\to \pi^+\pi^-\pi^-$ by the LHCb also differ by a factor of seven! This together with the theoretical predictions of $\B(B^-\to f_2(1270)\pi^-)$ leads us to conjecture that the $f_2(1270)$ contribution to $B^-\to K^+K^-\pi^-$ is largely overestimated experimentally . This needs to be clarified in the Run II experiment.
    Including $1/m_b$ power corrections from penguin annihilation inferred from QCDF, a sizable  \CP asymmetry of 32\% in the $f_2(1270)$ component are in accordance with the LHCb measurement.

\item A fraction of 5\% for the $\rho(1450)$ component in $B^-\to\pi^+\pi^-\pi^-$ is in accordance with the theoretical expectation. However, a large fraction of 30\% in $B^-\to K^+K^-\pi^-$ is entirely unexpected. If this feature is confirmed in the future, it is likely that the broad vector resonance
    $\rho(1450)$ may play the role of the s-called $f_X(1500)$ broad resonance observed in $B\to KKK$ and $K\ov K\pi$ decays.

\item The contribution of $K_0^*(1430)^0$ to $B^-\to K^+K^-\pi^-$ was found to be too large by a factor of 3 when confronted with experiment. The current theoretical predictions based on both QCDF and pQCD for $\B(B^-\to K_0^*(1430)^0 K^-)$ are also too large compared to experiment. This issue needs to be resolved.

\item  By varying the relative phase between $A_\sigma$ and $A_{\rm NR}^{\pi^+\pi^+}$, we find that a large and negative \CP asymmetry of $-66\%$ through the $S$-wave $\pi^+\pi^-\to K^+K^-$ rescattering can be accommodated at $\phi_\sigma\approx 134^\circ$. However, the predicted branching fraction is less than the LHCb value by a factor of 4!
    This is ascribed to the fact that one should use Eq. (\ref{eq:FSI}) to describe $\pi\pi\leftrightarrow K\ov K$ final-state rescattering. By the same token, the branching fraction of the rescattering contribution to $B^-\to \pi^+\pi^-\pi^-$ also seems to be overestimated experimentally by a factor of 4.

\item Using the QCDF expression of the $B^-\to \sigma\pi^-$ amplitude  to compute $B^-\to \sigma\pi^-\to\pi^+\pi^-\pi^-$, the resultant \CP violation of 15\% and branching fraction agree with experiment.

\item  \CP asymmetry for the dominant quasi-two-body decay mode $B^-\to\rho^0\pi^-$ was found by the LHCb to be consistent with zero in all three $S$-wave models. In the QCD factorization approach, the $1/m_b$ power corrections, namely penguin annihilation and hard spectator interactions, contribute destructively to $\acp(B^-\to\rho^0\pi^-)$ to render it consistent with zero.

\item While \CP violation in $B^-\to\rho^0\pi^-$ is consistent with zero, a significant \CP asymmetry has been seen  in the $\rho^0(770)$ region where the data are separated by the sign of the value of $\cos\theta$ with $\theta$ being the angle between the pions with the same-sign charge. Considering the low $\pi^+\pi^-$ invariant mass region of the $B^+\to \pi^+\pi^+\pi^-$ Dalitz plot of \CP asymmetries divided into four zones as depicted in Fig. \ref{fig:Dalitz}, we have predicated the sign of \CP violation in each zone correctly which arises from the interference between the $\rho(770)$ and $\sigma$ as well as the nonresonant background.

\end{itemize}

\vskip 2.0cm \acknowledgments

We are very grateful to Zhi-Tian Zou for helpful discussions.
This research was supported in part by the Ministry of Science and Technology of R.O.C. under Grant No. 106-2112-M-033-004-MY3.

\appendix

\section{Input parameters}

Many of the input parameters for the decay constants of pseudoscalar and vector mesons and form factors for $B\to P,V$ transitions can be found in \cite{CC:Bud} where uncertainties in form factors are shown. The reader is referred to \cite{Cheng:scalar} for decay constants and form factors related to scalar mesons. For reader's convenience, we list the scalar decay constants relevant to this work
\be
\bar f_{f_0}=460, \quad \bar f_\sigma^u=350, \quad \bar f_{K_0^*(1430)}=550\,,
\en
defined at $\mu=1$ GeV and expressed in units of MeV. The
vector decay constant of $K_0^*(1430)$ is related to the scalar one via
\be
f_{K_0^*}={m_s(\mu)-m_q(\mu)\over m_{K^*_0}}\bar f_{K_0^*}.
\en

The form factors used in this work are
\be
&& F_0^{B\pi}(0)=0.25\pm0.03, \qquad\quad~  F_0^{BK}(0)=0.35\pm0.04,  \non \\
&& A_0^{B\rho}(0)=0.303\pm0.029, \qquad~ A_0^{B\omega}(0)=0.281\pm0.030,  \non \\
&& A_0^{Bf_2}(0)=0.13\pm0.02,  \qquad\quad F^{B\sigma^u}(0)=0.25\pm0.02,  \\
&& A_2^{B\rho}(0)=0.221\pm0.023, \qquad~ A_2^{B\omega}(0)=0.198\pm0.023.  \non
\en
The $B\to f_2(1270)$ transition form factor taken from \cite{Cheng:TP} is evaluated using large energy effective theory, while
the form factors for $B\to V$ transition are from \cite{Ball:2004rg}.
There is an updated light-cone sum-rule analysis of
$B\to V$ transition form factors in \cite{Straub:2015ica} in which one has
\be
A_0^{B\rho}(0)=0.356\pm0.042, \qquad A_0^{B\omega}(0)=0.328\pm0.048\,.
\en
However, we will not use this new analysis in this study for two reasons. First, it will lead to too large $B^-\to \rho^0\pi^-$ and $B^-\to\omega\pi^-$ rates compared to experiment.  Second, the parameters $(\rho_A,\phi_A)$ and $(\rho_H,\phi_H)$, which govern $1/m_b$ power corrections from penguin annihilation and hard spectator interactions, respectively, have been extracted from the data
using $B\to V$ from factors given by \cite{Ball:2004rg}, see Appendix B below.

Note that for the $\sigma$ meson, the Clebsch-Gordon coefficient $1/\sqrt{2}$ is already included in $\bar f_\sigma^u$ and $F_0^{B\sigma^u}$. For the $f_0(980)$, one needs to multiple a factor of $\sin\theta/\sqrt{2}$ to get its decay constant and form factor, for example,
$\bar f_{f_0(980)}^u=\bar f_{f_0(980)}\sin\theta/\sqrt{2}$ with the mixing angle $\theta\approx 20^\circ$.

For the CKM matrix elements, we use the updated Wolfenstein parameters
$A=0.8235$, $\lambda=0.224837$, $\bar \rho=0.1569$ and $\bar
\eta=0.3499$ \cite{CKMfitter}. The corresponding CKM angles are
$\sin2\beta=0.7083^{+0.0127}_{-0.0098}$ and
$\gamma=(65.80^{+0.94}_{-1.29})^\circ$ \cite{CKMfitter}.

Among the quarks, the strange quark gives the major theoretical uncertainty to
the decay amplitude. Hence, we will only consider the uncertainty in the
strange quark mass given by $m_s(2\,{\rm GeV})=92.0\pm1.1$ MeV \cite{Aoki:2019cca}.

\section{Flavor operators}
In our previous works \cite{CCS:nonres,Cheng:2013dua,Cheng:2016shb}, we have employed the values of the flavor operators $a_i^p$ given in \cite{CCS:nonres} at the renormalization scale $\mu=\ov m_b/2=2.1$~GeV. Since then, there is a substantial progress in the determination of $1/m_b$ power corrections to $a_i^p$. In the QCD factorization approach, flavor operators have the expressions \cite{BBNS,BN}
 \be \label{eq:ai}
  a_i^{p}(M_1M_2) =
 \left(c_i+{c_{i\pm1}\over N_c}\right)N_i(M_2)
  + {c_{i\pm1}\over N_c}\,{C_F\alpha_s\over
 4\pi}\Big[V_i(M_2)+{4\pi^2\over N_c}H_i(M_1M_2)\Big]+P_i^{p}(M_2),
 \en
where $i=1,\cdots,10$,  the upper (lower) signs apply when $i$ is
odd (even), $c_i$ are the Wilson coefficients,
$C_F=(N_c^2-1)/(2N_c)$ with $N_c=3$, $M_2$ is the emitted meson
and $M_1$ shares the same spectator quark with the $B$ meson. The
quantities $V_i(M_2)$ account for vertex corrections,
$H_i(M_1M_2)$ for hard spectator interactions with a hard gluon
exchange between the emitted meson and the spectator quark of the
$B$ meson and $P_i^p(M_2)$ for penguin contractions.

In the QCD factorization approach, there are two kinds of $1/m_b$ corrections: penguin annihilation to the penguin amplitude and hard spectator interactions to $a_2$:
\be \label{eq:P}
P&=& P_{\rm SD}+1/m_b~{\rm corrections} \non \\
&\propto & [\lambda_u(a_4^u+r_\chi^P a_6^u)+\lambda_c(a_4^c+r_\chi^P a_6^c)] +
\underbrace{\lambda_u\beta_3^u+\lambda_c\beta_3^c}_{\rm penguin~annihilation},
\en
and
\be \label{eq:a2}
a_2(M_1M_2) = c_2+{c_1\over N_c} + {c_1\over N_c}\,{C_F\alpha_s\over
 4\pi}\Big[V_2(M_2)+{4\pi^2\over N_c}H_2(M_1M_2)\Big].
\en
Power corrections in QCDF often involve endpoint divergences. We shall follow \cite{BBNS} to model the endpoint divergence $X\equiv\int^1_0 dx/(1-x)$ in the penguin annihilation and hard spectator scattering diagrams as
\be \label{eq:XA}
 X_A^{i,f} = \ln\left({m_B\over \Lambda_h}\right)(1+\rho_A^{i,f} e^{i\phi_A^{i,f}}), \qquad
 X_H = \ln\left({m_B\over \Lambda_h}\right)(1+\rho_H e^{i\phi_H}),
\en
with $\Lambda_h$ being a typical hadronic scale of 0.5 GeV,
where the superscripts `$i$' and `$f$' refer to gluon emission from the initial and final-state quarks, respectively. A fit of  the four parameters $(\rho_A^{i,f},\phi_A^{i,f})$ with the first order approximation of $\rho_H\approx \rho_A^i$ and $\phi_H\approx \phi_A^i$ to the $B\to PP$ and $PV$ data yields \cite{Sun:2014tfa,Chang:2014yma}
\be
(\rho_A^i,\rho_A^f)_{_{P\!P}}=(2.98^{+1.12}_{-0.86}, 1.18^{+0.20}_{-0.23}), \qquad
(\phi_A^i,\phi_A^f)_{_{P\!P}}=(-105^{+34}_{-24}, -40^{+11}_{-8})^\circ,
\en
and
\be \label{eq:rhoPV}
(\rho_A^i,\rho_A^f)_{_{PV}}=(2.87^{+0.66}_{-1.95}, 0.91^{+0.12}_{-0.13}), \qquad
(\phi_A^i,\phi_A^f)_{_{PV}}=(-145^{+14}_{-21}, -37^{+10}_{-9})^\circ.
\en

In general, the difference between $a_i^p(M_2M_1)$ and $a_i^p(M_1M_2)$ is small for the quasi-two-body decays $B\to PV$ except for $a_{6,8}^p$. Using Eq. (\ref{eq:rhoPV}) as an input for $1/m_b$ power corrections and taking the averages of $a_i^p(PV)$ and $a_i^p(VP)$ (except for $a_{6,8}^p$), we have
 \be \label{eq:ai_value}
 && a_1\approx0.988\pm0.102 i,\quad a_2\approx 0.183-0.348i, \quad a_3\approx -0.0023+0.0174i, \quad a_5\approx
 0.00644-0.0231i,  \non \\
 && a_4^u\approx -0.025-0.021i, \quad a_4^c\approx
 -0.030-0.012i,\quad
 a_6^u\approx -0.042-0.014i, \quad a_6^c\approx -0.045-0.005i,
 \non\\
 &&a_7\approx (-0.5+2.7i)\times 10^{-4},\quad a_8^u\approx (5.2-1.0i)\times
 10^{-4},\quad
 a_8^c\approx (5.0-0.5i)\times
 10^{-4},   \\
 && a_9\approx (-8.9-0.9i)\times 10^{-3},\quad
 a_{10}^u \approx (-1.45+ 3.12 i)\times10^{-3},\quad
 a_{10}^c \approx (-1.51 + 3.17 i)\times10^{-3}, \non
 \en
at the renormalization scale $\mu=\ov m_b(\ov m_b)=4.18$~GeV, where the values of $a_{6,8}^p$ are for $M_1M_2=VP$. For $M_1M_2=PV$ we should use
\be \label{eq:ai68}
 && a_6^u(PV)\approx -0.010-0.015i, \qquad\quad a_6^c(PV)\approx -0.013-0.006i,
 \non\\
 && a_8^u(PV)\approx -(8.9+8.5i)\times 10^{-5},\quad
 a_8^c(PV)\approx -(10.7+3.7i)\times 10^{-4}.
 \en
There are two different sources for the strong phases of $a_i^p$: (i) vertex corrections, hard spectator interactions and penguin contractions which are perturbatively calculable in the QCD factorization approach \cite{BBNS} and (ii) $1/m_b$ power corrections.

Note that the parameter $N_i(M)$ in Eq. (\ref{eq:ai}) vanishes if $M$ is a tensor meson or a vector meson with $i=6,8$, or a neutral scalar meson such as $\sigma,f_0$ and $a_0^0$ with $i\neq 6,8$. Otherwise, it is equal to one.
Consequently,  the flavor operators given in Eqs. (\ref{eq:ai_value}) and (\ref{eq:ai68}) are not applicable to the quasi-two-body decays $B\to SP$ ($S=\sigma$ or $f_0$) for two reasons. First, $N_i(\sigma)=0$ means that $a_i^p(P\sigma)$ do not receive factorizable contributions
except for $i=6,8$. Second, light-cone distribution amplitudes (LCDAs) of scalar and pseudoscalar mesons have different behavior. While the symmetric pion LCDA peaks at $x=1/2$, the antisymmetric LCDA of the light scalar such as $\sigma$ peaks at $x=0.25$ and 0.75. As a result, $a_i^p(\sigma P)$ and $a_i^p(P\sigma)$ can be quite different except for $a_{6,8}^p$.
As an example, numerical values of the flavor operators $a_i^p(M_1M_2)$ for $M_1M_2=\sigma\pi$ and $\pi\sigma$ are shown in Table \ref{tab:aiSP}. We see that, for instance,
$a_1(\pi\sigma)=0.015-0.004i$ is very different from $a_1(\sigma\pi)=0.95+0.014i$. In practice, we also use the same set of flavor operators to work out $B\to f_0(980)\pi$ decays.


\begin{table}[t]
\caption{Numerical values of the flavor operators $a_i^p(M_1M_2)$ for $M_1M_2=\sigma\pi$ and $\pi\sigma$ at the scale $\mu=\ov m_b(\ov m_b)=4.18$ GeV \cite{Cheng:2020hyj}. In this work we use the same set of flavor operators to work out $B\to f_0(980)\pi$ decays.}
\label{tab:aiSP}
\begin{center}
\begin{tabular}{ l c c | l r r} \hline \hline
 $a_i^p$ & ~~$\sigma\pi$~~ & ~~~$\pi\sigma$~~~ & ~~$a_i^p$~~  & ~~$\sigma\pi$~~ & ~~$\pi\sigma$ \\
\hline
 $a_1$ & ~~~$0.95+0.014i$~~~ & ~~$0.015-0.004i$~~ & ~~$a_6^c$ & $-0.045-0.005i$ & $-0.045-0.005i$  \\
 $a_2$ & $0.33-0.080i$ & $-0.056+0.024i$ &  ~~$a_7$ & $(-1.8+0.3i)10^{-4}$ & $(-4.2+1.0i)10^{-5}$ \\
 $a_3$ & $-0.009+0.003i$ & $0.0026-0.0008i$ & ~~$a_8^u$ & $(4.8-1.0i)10^{-4}$ & $(4.8-1.0i)10^{-4}$  \\
 $a_4^u$ & $-0.022-0.015i$ & $0.062-0.013i$ & ~~$a_8^c$ & $(4.6-0.5i)10^{-4}$  &  $(4.6-0.5i)10^{-4}$\\
 $a_4^c$ & $-0.027-0.006i$ & $-0.012-0.007i$ & ~~$a_9$  & $(-8.6-0.1i)10^{-3}$ & $(-1.3+0.4i)10^{-4}$\\
 $a_5$ & $0.0158-0.003i$ & $0.0035-0.0009i$ & ~~$a_{10}^u$ & $(-2.6+0.6i)10^{-3}$  & $(8.7-3.1i)10^{-4}$ \\
 $a_6^u$ & $-0.042-0.014i$ & $-0.042-0.014i$  & ~~$a_{10}^c$ & $(-2.6+0.7i)10^{-3}$ & $(4.6-2.8i)10^{-4}$ \\
\hline \hline
\end{tabular}
\end{center}
\end{table}

\section{Final State Interactions}

Since there are some confusions in the literature concerning the rescattering formula,
we believe that it will be useful to go through the relevant derivations.
Our discussion follows Refs. \cite{Chua:2007cm,tdlee} closely.
The weak Hamiltonian is given by $H_{\rm W}=\sum_q\lambda_q O_q$,
where $\lambda_q$ are $V_{qb} V^*_{qd}$ and $O_q$ are four-quark operators with Wilson coefficients included.
From the time reversal
invariance of $O_q$, it follows that
\begin{eqnarray}
(\langle i;{\rm out}| O_q|\overline B\rangle)^*
 =\left(\langle i;{\rm out}|\right)^*U^\dagger_T U_T O_{q}^* U^\dagger_T U_T |\overline B\rangle^*
 = \langle i;{\rm in}|O_q|\overline B\rangle,
 \en
which can be expressed as
 \be
(\langle i;{\rm out}| O_q|\overline B\rangle)^*
 = \sum_k\langle i;{\rm in}|\,k;{\rm out}\rangle
          \langle k;{\rm out}|O_q|\overline B\rangle
 = \sum_k {\cal S}^\dagger_{ik} \langle k;{\rm out}|O_q|\overline B\rangle,
 \label{eq:timerev}
\end{eqnarray}
where ${\cal S}_{ik}\equiv\langle i;{\rm out}|\,k;{\rm in}\rangle$ denotes the
strong interaction $S$-matrix element.
Note that we have used $U_T(|{\rm out\,(in)}\rangle)^*=|{\rm in\,(out)}\rangle$
to fix the phase convention, which also leads to
\be
{\cal S}^*_{ij}=(\langle i;{\rm out}|)^* U^\dagger_T U_T(|j;{\rm in}\rangle)^*
=\langle i;{\rm in}|\,j;{\rm out}\rangle={\cal S}^*_{ji}.
\label{eq: Sij=Sji}
\en
From the following identity
\be
\sum_k{\cal S}^\dagger_{ik} {\cal S}^{1/2}_{kj}=({\cal S}^{1/2})^\dagger_{ij}=({\cal S}^{1/2})^*_{ji}=({\cal S}^{1/2})^*_{ij},
\en
where use of Eq. (\ref{eq: Sij=Sji}) has been made,
it is clear that the solution of
Eq. (\ref{eq:timerev}) is simply \cite{Suzuki}
\begin{equation}
 \langle i; {\rm out}|O_q|\overline B\rangle=\sum_j {\cal S}^{1/2}_{ij} {\cal A}^{q}_{0j},
 \label{eq:FSIsolution}
\end{equation}
where ${\cal A}^{q}_{0j}$ is a real amplitude.
The weak decay amplitude picks up strong scattering
phases \cite{Watson:1952ji} and finally we have \cite{Chua:2007cm}
 \be
\langle i; {\rm out}|H_{\rm W}|\overline B\rangle
           =\sum_q\langle i;{\rm out}|\lambda_q O_q|\overline B\rangle
           =\sum_{q,j} {\cal S}^{1/2}_{ij} (\lambda_q {\cal A}^{q}_{0j})
           =\sum_j {\cal S}^{1/2}_{ij} {\cal A}_{0j},
 \label{eq:FSIsolutionfull}
 \en
where we have defined ${\cal A}_0\equiv \sum_q \lambda_q {\cal A}^q_0$ and, consequently, it is free of any strong phase.

It will be useful to give an equivalent expression to the above results in terms of time evolution operator~\cite{tdlee}.
It is well known that the so-called `in' and `out' states can be expressed as
\be
|i; {\rm in}\ra=\lim_{T\to\infty} U_I(0,-T)|i; {\rm free}\ra,
\quad
|i; {\rm out}\ra=\lim_{T\to\infty} U_I(0,T)|i; {\rm free}\ra,
\en
with $U_I(t_2, t_1)$ the time evolution operator in the interaction picture given by
\be
U_I(t_2, t_1)= e^{i H_0 t_2} e^{-i H( t_2-t_1)} e^{-i H_0 t_1}=e^{i H_0 t_2} U( t_2,t_1) e^{-i H_0 t_1},
\label{eq: UI}
\en
where $H_0$ is the free Hamiltonian and $H$ is the full strong Hamiltonian. The time evolution operator satisfies $U^\dagger_T U^*_I(t_2,t_1) U_T=U_I(-t_2, -t_1)$ and $U^\dagger_I (t_2,t_1)=U_I(t_1,t_2)$, as $H_0$ and $H$ are time-invariant and hermitian.

The amplitude $\langle i; {\rm out}|O_q|\overline B\rangle$ can now be expressed as
\be
\langle i; {\rm out}|O_q|\overline B\rangle=\lim_{T\to \infty}\la i; {\rm free}|U_I(T,0) O_q|\overline B\ra,
\en
and the previous derivations can all be brought through parallelly with the help of $U_T(|i;{\rm free}\ra)^*=|i;{\rm free}\ra$ matching the phase convention and the time invariant properties of $H_0$ and $H$.
Indeed, from
\be
U^\dagger_T U^*_I(T,0) U_T=U_I(-T,0)=U_I(-T,T)U_I(T,0)=U^\dagger_I(T,-T) U(T,0),
\en
and
\be
{\cal S}_{ij}&\equiv& \langle i;{\rm out}|\,j;{\rm in}\rangle=\lim_{T\to\infty}\la i;{\rm free}| U_I(T,-T)|j;{\rm free}\ra,
\en
we have
\be
(\lim_{T\to \infty}\la i; {\rm free}| U_I(T,0) O_q|\overline B\ra)^*
&=&\sum_k\lim_{T\to \infty}\la i; {\rm free}|U^\dagger_I(T,-T)|k,{\rm free}\ra
                                       \la k; {\rm free}| U_I(T,0) O_q|\overline B\ra,
\non\\
\en
which is equivalent to Eq.~(\ref{eq:timerev}).
Furthermore, using Eq. (\ref{eq: UI}) and the fact that $|i, {\rm free}\ra$ and $|j, {\rm free}\ra$ are degenerate eigenstates of $H_0$, we are led to
\be
\la i;{\rm free}| U_I(T,0)|j;{\rm free}\ra
=\la i;{\rm free}| U_I(0,-T)|j;{\rm free}\ra
=\la i;{\rm free}| U_I(T/2,-T/2)|j;{\rm free}\ra,
\en
which justifies the following definition,
\be
{\cal S}^{1/2}_{ij}&\equiv& \lim_{T\to\infty}\la i;{\rm free}| U_I(T,0)|j;{\rm free}\ra,
\label{eq: S1/2}
\en
and, consequently, with ${\cal A}^{q}_{0j}\equiv\la j;{\rm free}| O_q|\overline B\ra$, we obtain
\be
\langle i; {\rm out}|Q_q|\overline B\rangle
 =\lim_{T\to\infty}\la i;{\rm free}| U_I(T,0) O_q|\overline B\rangle
                      =\sum_j {\cal S}^{1/2}_{ij} \la j;{\rm free}|O_q|\overline B\ra
                      =\sum_j {\cal S}^{1/2}_{ij} {\cal A}^{q}_{0j},
\en
which corresponds to Eq. (\ref{eq:FSIsolution}), and Eq. (\ref{eq:FSIsolutionfull}) follows accordingly.
Note that ${\cal A}^{q}_{0j}=({\cal A}^{q}_{0j})^*$ is a consequence of the phase convention and the time invariant property of $O_q$.

It is useful to express Eq. (\ref{eq:FSIsolutionfull}) in term of the full time-evolution operator,
\be
\langle i; {\rm out}|H_{\rm W}|\overline B\rangle
 =\lim_{T\to\infty}\la i;{\rm free}| e^{i H_0 T} U(T,0) H_{\rm W}|\overline B\rangle,
 \en
which can be decomposed into (with $\tau\gsim 0$)
\be
\langle i; {\rm out}|H_{\rm W}|\overline B\rangle
 =\sum_j \lim_{T\to\infty}\la i;{\rm free}| e^{i H_0 T} U(T,\tau) e^{-i H_0 \tau} |j,{\rm free}\ra
 \la j; {\rm free}| e^{i H_0 \tau}  U(\tau, 0) H_{\rm W}|\overline B\rangle.
 \en
The above expression clearly shows the time evolution nature of rescattering~\cite{wshou} and the rescattering of $\pi\pi\to K\ov K$ is considered to happen at a much later stage of time-evolution contained in
$\la i;{\rm free}| e^{i H_0 T} U(T,\tau) e^{-i H_0 \tau} |j,{\rm free}\ra$,
while all the violent and rapid interactions have already happened and are contained in
$ \la j; {\rm free}| e^{i H_0 \tau}  U(\tau, 0) H_{\rm W}|\overline B\ra$.



\end{document}